\documentclass[journal]{IEEEtran}
\usepackage{amsmath}
\usepackage{amsfonts}
\usepackage{amssymb}
\usepackage{epsf}
\usepackage{cite}
\usepackage{balance}
\usepackage{fancyhdr}
\usepackage{graphicx}
\usepackage{subfigure}
\usepackage[blocks]{authblk}
\usepackage[stable]{footmisc}
\usepackage{hyperref}
\usepackage{cleveref}
\usepackage{fancyhdr}

\makeatletter
\def\ps@IEEEtitlepagestyle{%
  \def\@oddfoot{\mycopyrightnotice}%
  \def\@evenfoot{}%
}
\def\mycopyrightnotice{%
  {\footnotesize \copyright 2015 IEEE. Personal use of this material is permitted. Permission from IEEE must be obtained for all other uses, in any current or future media\hfill}
  \gdef\mycopyrightnotice{}
}

\newcounter{subeq}
\renewcommand{\thesubeq}{\theequation\alph{subeq}}
\newcommand{\newsubeqblock}{\setcounter{subeq}{0}\refstepcounter{equation}}
\newcommand{\mysubeq}{\refstepcounter{subeq}\tag{\thesubeq}}
\date{}
\begin{document}

\title{A Novel Statistical Channel Model for Turbulence-Induced Fading in Free-Space Optical Systems}

\author{Mohammadreza~A. Kashani,~\IEEEmembership{Student Member,~IEEE,}
        Murat~Uysal,~\IEEEmembership{Senior Member,~IEEE,}
        and~Mohsen~Kavehrad,~\IEEEmembership{Fellow,~IEEE}

\thanks{Mohammadreza A. Kashani and Mohsen Kavehrad are with Pennsylvania State University, University Park, PA 16802 (email:  mza159@psu.edu, mkavehrad@psu.edu).}
\thanks{Murat Uysal is with Department of Electrical and Electronics Engineering, \"{O}zye\v{g}in University, Istanbul, Turkey, 34794 (email: murat.uysal@ozyegin.edu.tr).}
\thanks{This paper was presented in part at the IEEE international Conference on Transparent Optical Networks (ICTON), Cartagena, Spain, June 2013. The work of M. Uysal is supported by TUBITAK Research Grant 111E143.}}
\maketitle

\begin{abstract}
In this paper, we propose a new probability distribution function which accurately describes turbulence-induced fading under a wide range of turbulence conditions. The proposed model, termed Double Generalized Gamma (Double GG), is based on a doubly stochastic theory of scintillation and developed via the product of two Generalized Gamma (GG) distributions. The proposed Double GG distribution generalizes many existing turbulence channel models and provides an excellent fit to the published plane and spherical waves simulation data. Using this new statistical channel model, we derive closed form expressions for the outage probability and the average bit error as well as corresponding asymptotic expressions of free-space optical communication systems over turbulence channels. We demonstrate that our derived expressions cover many existing results in the literature earlier reported for Gamma-Gamma, Double-Weibull and K channels as special cases.\end{abstract}


\section{Introduction}\label{INTRODUCTION}
\IEEEPARstart{F}{ree-space} optical (FSO) communication enables wireless connectivity through atmosphere using laser transmitters at infrared bands. These systems provide high data rates comparable to fiber optics while they offer much more flexibility in (re)deployment. Since they operate in unregulated spectrum, no licensing fee is required making them also a cost-effective solution \cite{1,2,kashani2}. With their unique features and advantages, FSO systems have attracted attention initially as a “last mile” solution and can be used in a wide array of applications including cellular backhaul, inter-building connections in enterprise/campus environments, video surveillance/monitoring, fiber back-up, redundant link in disaster recovery and relief efforts among others.

A major performance limiting factor in FSO systems is atmospheric turbulence-induced fading (also called as scintillation) \cite{3}. Inhomogenities in the temperature and the pressure of the atmosphere result in variations of the refractive index and cause atmospheric turbulence. This manifests itself as random fluctuations in the received signal and severely degrades the FSO system performance particularly over long ranges.

In the literature, several statistical models have been proposed in an effort to model this random phenomenon. Historically, log-normal distribution has been the most widely used model for the probability density function (pdf) of the random irradiance over atmospheric channels \cite{4,5,6}. This pdf model is however only applicable to weak turbulence conditions. As the strength of turbulence increases, lognormal statistics exhibit large deviations compared to experimental data. Moreover, lognormal pdf underestimates the behavior in the tails as compared with measurement results. Since the calculation of detection probabilities for a communication system is primarily based on the tails of the pdf, underestimating this region significantly affects the accuracy of performance analysis.

In an effort to address the shortcomings of the lognormal distribution, other statistical models have been further proposed to describe atmospheric turbulence channels under a wide range of turbulence conditions. These include the Negative Exponential/Gamma model (also known widely as the K channel) \cite{7}, I-K distribution \cite{8}, log-normal Rician channel (also known as Beckman) \cite{9}, Gamma-Gamma \cite{10}, M distribution \cite{11} and Double-Weibull \cite{12}. Particularly worth mentioning is the Gamma-Gamma model \cite{10}, \cite{13} which has been widely used in the literature for the performance analysis of FSO systems, see e.g., \cite{14,15}, along with the log-normal model. This model builds upon a two-parameter distribution and considers irradiance fluctuations as the product of small-scale and large-scale fluctuations, where both are governed by independent gamma distributions. In a more recent work by Chatzidiamantis \emph{et al.} in \cite{12}, the Double-Weibull distribution was proposed as a new model for atmospheric turbulence channels. Similar to the Gamma-Gamma model, it is based on the theory of doubly stochastic scintillation and considers irradiance fluctuations as the product of small-scale and large-scale fluctuations which are both Weibull distributed. It is shown in \cite{12} that Double-Weibull is more accurate than the Gamma-Gamma particularly for the cases of moderate and strong turbulence.

In this paper, we propose a new and unifying statistical model, named Double Generalized Gamma (Double GG), for the irradiance fluctuations. The proposed model is valid under all range of turbulence conditions (weak to strong) and contains most of the existing statistical models for the irradiance fluctuations in the literature as special cases. Furthermore, we provide comparison of the proposed model with Gamma-Gamma and Double-Weibull models. For this purpose, we use the set of simulation data from \cite{16,17} for plane and spherical waves\footnote{The simulation data in \cite{16,17} was obtained through phase screen approach which consists of approximating a three-dimensional random medium as a collection of equally spaced, two-dimensional, random phase screens that are transverse to the direction of wave propagation. In \cite{16}, it was discussed in detail that such a numerical simulation approach contains all the essential physics for accurately predicting the pdf of irradiance (or, equivalently, log-normal irradiance), and shown that the simulation results provide an excellent match to known experimental measurements reported in \cite{exp} for both plane and spherical waves. The same set of simulation data was also used in \cite{10} and \cite{12} which respectively introduced Gamma-Gamma and Double-Weibull distributions as turbulence channel models.}. Our model demonstrates an excellent match to the simulation data and is clearly superior over the other two models which show discrepancy from the simulation data in some cases. In the second part of the paper, we use this new channel model to derive closed form expressions for the BER and the outage probability of single-input single-output (SISO) and single-input multiple-output (SIMO) FSO systems with intensity modulation and direct detection (IM/DD). Our performance results can be seen as a generalization of the results in \cite{18,19,20,21}.

The rest of the paper is organized as follows: In Section II, we propose Double GG distribution to characterize turbulence-induced fading. In Section III, we confirm the accuracy of our model through comparisons with simulation data for plane and spherical waves. In Section IV, we present the derivation of bit error rate (BER) and outage probability for SISO FSO system over Double GG channel. In Section V, we present BER expressions for FSO links with multiple receiver apertures. Finally, Section VI concludes the paper.
\section{Double GG Distribution}\label{dgg}
The irradiance of the received optical wave can be modeled as \cite{10}, \cite{12} $I={{I}_{x}}{{I}_{y}}$, where ${{I}_{x}}$ and ${{I}_{y}}$ are statistically independent random processes arising respectively from large-scale and small scale turbulent eddies. We assume that both large-scale and small-scale irradiance fluctuations are governed by Generalized Gamma (GG) distributions [\citen{23}, Eq. (1)]. The pdfs of ${{I}_{x}}\sim GG\left( {{\gamma }_{1}},{{m}_{1}},{{\Omega }_{1}} \right)$  and ${{I}_{y}}\sim GG\left( {{\gamma }_{2}},{{m}_{2}},{{\Omega }_{2}} \right)$ are given by
\begin{equation}\label{eq1}
{{f}_{{{I}_{x}}}}\left( {{I}_{x}} \right)=\frac{{{\gamma }_{1}}I_{x}^{{{m}_{1}}{{\gamma }_{1}}-1}}{{{\left( {{{\Omega }_{1}}}/{{{m}_{1}}}\; \right)}^{{{m}_{1}}}}\Gamma \left( {{m}_{1}} \right)}\exp \left( -\frac{{{m}_{1}}}{{{\Omega }_{1}}}I_{x}^{{{\gamma }_{1}}} \right)
\end{equation}
\begin{equation}\label{eq2}
{{f}_{{{I}_{y}}}}\left( {{I}_{y}} \right)=\frac{{{\gamma }_{2}}I_{y}^{{{m}_{2}}{{\gamma }_{2}}-1}}{{{\left( {{{\Omega }_{2}}}/{{{m}_{2}}}\; \right)}^{{{m}_{2}}}}\Gamma \left( {{m}_{2}} \right)}\exp \left( -\frac{{{m}_{2}}}{{{\Omega }_{2}}}I_{y}^{{{\gamma }_{2}}} \right)
\end{equation}
where ${{\gamma }_{i}}>0$ , ${{m}_{i}}\ge 0.5$ and ${{\Omega }_{i}}$ $i=1,2$ are the GG parameters. The pdf of $I$ can be written as
\begin{equation}\label{eq3}
{{f}_{I}}\left( I \right)=\int\limits_{0}^{\infty }{{{f}_{{{I}_{x}}}}\left( I|{{I}_{y}} \right){{f}_{{{I}_{y}}}}\left( {{I}_{y}} \right)d{{I}_{y}}}
\end{equation}
where ${{f}_{{{I}_{x}}}}\left( I|{{I}_{y}} \right)$ is obtained as
\begin{equation}
{{f}_{{{I}_{x}}}}\left( I|{{I}_{y}} \right)=\frac{{{\gamma }_{1}}{{\left( I/{{I}_{y}} \right)}^{{{m}_{1}}{{\gamma }_{1}}-1}}}{{{I}_{y}}{{\left( {{{\Omega }_{1}}}/{{{m}_{1}}}\; \right)}^{{{m}_{1}}}}\Gamma \left( {{m}_{1}} \right)}\exp \left( -\frac{{{m}_{1}}}{{{\Omega }_{1}}}{{\left( \frac{I}{{{I}_{y}}} \right)}^{{{\gamma }_{1}}}} \right)
\end{equation}
The integration in (\ref{eq3}) yields
\begin{align}\label{eq4}
&{{f}_{I}}\left( I \right)=\frac{{{\gamma }_{2}}p{{p}^{{{m}_{2}}-1/2}}{{q}^{{{m}_{1}}-1/2}}{{\left( 2\pi  \right)}^{{1-\left( p+q \right)}/{2}\;}}{{I}^{-1}}}{\Gamma \left( {{m}_{1}} \right)\Gamma \left( {{m}_{2}} \right)}\\\nonumber
&\times G_{p+q,0}^{0,p+q}\left[ {{\left( \frac{{{\Omega }_{2}}}{{{I}^{{{\gamma }_{2}}}}} \right)}^{p}}\frac{{{p}^{p}}{{q}^{q}}\Omega _{1}^{q}}{m_{1}^{q}m_{2}^{p}}|\begin{matrix}
   \Delta \left( q:1-{{m}_{1}} \right),\Delta \left( p:1-{{m}_{2}} \right)  \\
   -  \\
\end{matrix} \right]
\end{align}
where $G_{p,q}^{m,n}\left[ . \right]$ is the Meijer’s G-function\footnote{Meijer’s G-function is a standard built-in function in mathematical software packages such as MATLAB, MAPLE and MATHEMATICA. If required, this function can be also expressed in terms of the generalized hypergeometric functions using [\citen{24}, Eqs.(9.303-304)].} defined in [\citen{24}, Eq.(9.301)], $p$ and $q$ are positive integer numbers that satisfy ${p}/{q}\;={{{\gamma }_{1}}}/{{{\gamma }_{2}}}\;$ and $\Delta (j;x)\triangleq {x}/{j}\;,...,{\left( x+j-1 \right)}/{j}\;$. We name this new distribution as Double GG. Employing [\citen{25}, Eq. (10)] and after some simplifications, the cumulative distribution function (cdf) of Double GG distribution can be obtained as
\begin{align}\label{eq5}
&{{F}_{I}}\left( I \right)=\frac{{{p}^{{{m}_{2}}-1/2}}{{q}^{{{m}_{1}}-1/2}}{{\left( 2\pi  \right)}^{{1-\left( p+q \right)}/{2}\;}}}{\Gamma \left( {{m}_{1}} \right)\Gamma \left( {{m}_{2}} \right)}\\\nonumber
&\times G_{1,p+q+1}^{p+q,1}\left[ {{\left( \frac{{{I}^{{{\gamma }_{2}}}}}{{{\Omega }_{2}}} \right)}^{p}}\frac{m_{1}^{q}m_{2}^{p}}{{{p}^{p}}{{q}^{q}}\Omega _{1}^{q}}|\begin{matrix}
   1  \\
   \Delta \left( q:{{m}_{1}} \right),\Delta \left( p:{{m}_{2}} \right),0  \\
\end{matrix} \right]
\end{align}
The distribution parameters ${{\gamma }_{i}}$ and ${{\Omega }_{i}}$ $i=1,2$ of the Double GG model can be identified using the first and second order moments of small and large scale irradiance fluctuations. The latter are directly tied to the atmospheric parameters. Without loss of generality, we assume $E\left( I \right)=1$ or equivalently $E\left( {{I}_{x}} \right)=1$ and $E\left( {{I}_{y}} \right)=1$. The second moment of irradiance is expressed as
\begin{equation}\label{eq6}
E\left( {{I}^{2}} \right)=E\left( I_{x}^{2} \right)E\left( I_{y}^{2} \right)=\left( 1+\sigma _{x}^{2} \right)\left( 1+\sigma _{y}^{2} \right)
\end{equation}
where $\sigma _{x}^{2}$ and $\sigma _{y}^{2}$ are respectively normalized variances of ${{I}_{x}}$ and ${{I}_{y}}$. The $n^\text{th}$ moment of ${{I}_{x}}$ (similarly${{I}_{y}}$) is given by
\begin{equation}\label{eq7}
E\left( I_{x}^{n} \right)={{\left( \frac{{{\Omega }_{1}}}{{{m}_{1}}} \right)}^{n/{{\gamma }_{1}}}}\frac{\Gamma \left( {{m}_{1}}+n/{{\gamma }_{1}} \right)}{\Gamma \left( {{m}_{1}} \right)}
\end{equation}
Therefore, by inserting the second order moment obtained from (\ref{eq7}) in (\ref{eq6}), and considering that $E\left( I \right)=1$, we have
\begin{align}\label{eq8a}
\newsubeqblock
\mysubeq &\sigma _{x}^{2}=\frac{\Gamma \left( {{m}_{1}}+{2}/{{{\gamma }_{1}}}\; \right)\Gamma \left( {{m}_{1}} \right)}{{{\Gamma }^{2}}\left( {{m}_{1}}+{1}/{{{\gamma }_{1}}}\; \right)}-1\\\label{eq8b}
\mysubeq &\sigma _{y}^{2}=\frac{\Gamma \left( {{m}_{2}}+{2}/{{{\gamma }_{2}}}\; \right)\Gamma \left( {{m}_{2}} \right)}{{{\Gamma }^{2}}\left( {{m}_{2}}+{1}/{{{\gamma }_{2}}}\; \right)}-1\\\label{eq9}
&{{\Omega }_{i}}={{\left( \frac{\Gamma \left( {{m}_{i}} \right)}{\Gamma \left( {{m}_{i}}+1/{{\gamma }_{i}} \right)} \right)}^{{{\gamma }_{i}}}}{{m}_{i}},~~i=1,2
\end{align}
where $m_i$ is a distribution shaping parameter and found using curve fitting on the simulated/measured channel data. Note that in (\ref{eq8a}) and (\ref{eq8b}), the variances of small- and large-scale ﬂuctuations (i.e., $\sigma _{x}^{2}$ and $\sigma _{y}^{2}$) are directly tied to the atmospheric conditions. Particularly, assuming a plane wave when inner scale effects are considered, the variances for the large- and the small-scale scintillations are given by [\citen{3}, Eqs. 9.46 and 9.55]
\begin{align}\nonumber
  & \sigma _{x}^{2}=\exp \left[ 0.16\sigma _{\text{Rytov}}^{2}{{\left( \frac{2.61{{\eta }_{l}}}{2.61+{{\eta }_{l}}+0.45\sigma _{\text{Rytov}}^{2}\eta _{l}^{7/6}} \right)}^{7/6}} \right.\\\nonumber
 & \times \left( 1+1.753{{\left( \frac{2.61}{2.61+{{\eta }_{l}}+0.45\sigma _{\text{Rytov}}^{2}\eta _{l}^{7/6}} \right)}^{1/2}} \right.\\\label{R1}
 &-\left.\left.0.252{{\left( \frac{2.61}{2.61+{{\eta }_{l}}+0.45\sigma _{\text{Rytov}}^{2}\eta _{l}^{7/6}} \right)}^{7/12}} \right) \right]-1
\end{align}
\begin{equation}\label{R2}
\sigma _{y}^{2}\cong \exp \left[ \frac{0.51\sigma _{\text{Rytov}}^{2}}{{{\left( 1+0.69\sigma _{\text{Rytov}}^{12/5} \right)}^{5/6}}} \right]-1
\end{equation}
where ${{\eta }_{l}}=10.89\left( {{R}_{0}}/{{l}_{0}} \right)$, and ${{R}_{0}}/{{l}_{0}}$ denotes the ratio of Fresnel zone to finite inner scale.
\begin{figure}
\centering
\includegraphics[width = 8cm, height = 7.5cm]{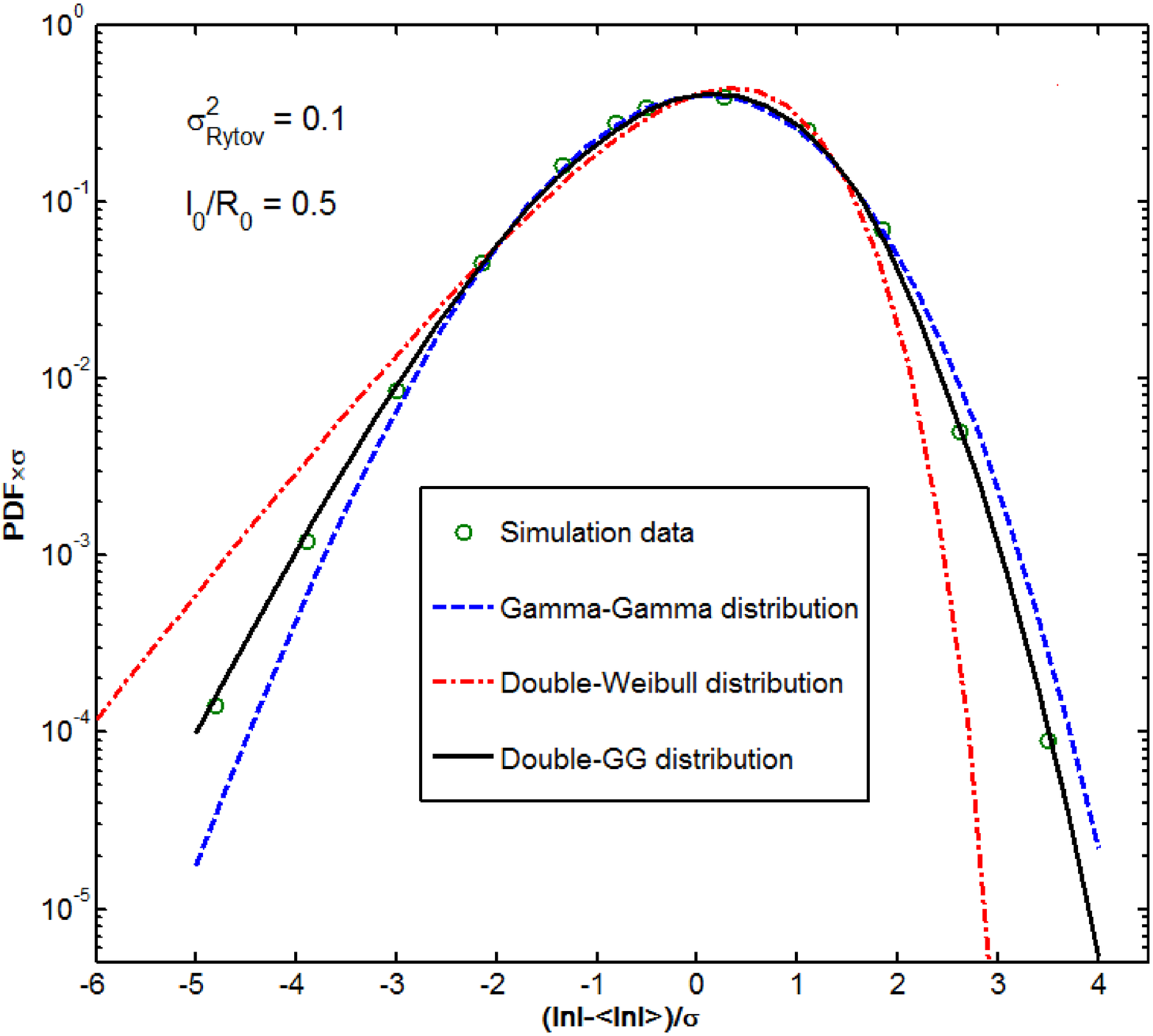}
\caption{Pdfs of the scaled log-irradiance for a plane wave assuming weak irradiance fluctuations.}
\end{figure}

For spherical waves in the absence of inner scale, $\sigma _{x}^{2}$ and $\sigma _{y}^{2}$ are given by [\citen{3}, Eqs. 9.63 and 9.70]
\begin{align}\label{R3}
&\sigma _{x}^{2}\cong \exp \left[ \frac{0.49\beta _{\text{0}}^{2}}{{{\left( 1+0.56\beta _{\text{0}}^{12/5} \right)}^{7/6}}} \right]-1\\\label{R4}
&\sigma _{y}^{2}\cong \exp \left[ \frac{0.51\beta _{\text{0}}^{2}}{{{\left( 1+0.69\beta _{0}^{12/5} \right)}^{5/6}}} \right]-1
\end{align}
where $\beta _{0}^{2}$ is the Rytov scintillation index of a spherical wave given by
\begin{equation}\label{R5}
\beta _{0}^{2}=\sigma _{\text{Rytov}}^{2}/{{\tilde{\sigma }}^{2}}\left( {{l}_{0}}/{{R}_{0}} \right)
\end{equation}
In (\ref{R5}), ${{\tilde{\sigma }}^{2}}\left( {{l}_{0}}/{{R}_{0}} \right)$ is defined as
\begin{align}\nonumber
& {{{\tilde{\sigma }}}^{2}}\left( {{l}_{0}}/{{R}_{0}} \right)\cong 3.86\left[ {{\left( 1+9/\eta _{l}^{2} \right)}^{11/12}}\left( \sin \left( \frac{11}{6}{{\tan }^{-1}}\frac{{{\eta }_{l}}}{3} \right)\right. \right.\\\nonumber
&+\frac{2.61}{{{\left( 9+\eta _{l}^{2} \right)}^{1/4}}}\sin \left( \frac{4}{3}{{\tan }^{-1}}\frac{{{\eta }_{l}}}{3} \right)\\
& \left.\left. -\frac{0.518}{{{\left( 9+\eta _{l}^{2} \right)}^{7/24}}}\sin \left( \frac{5}{4}{{\tan }^{-1}}\frac{{{\eta }_{l}}}{3} \right) \right)-8.75\eta _{l}^{-5/6} \right]
\end{align}
\begin{figure}
\centering
\includegraphics[width = 8cm, height = 7.5cm]{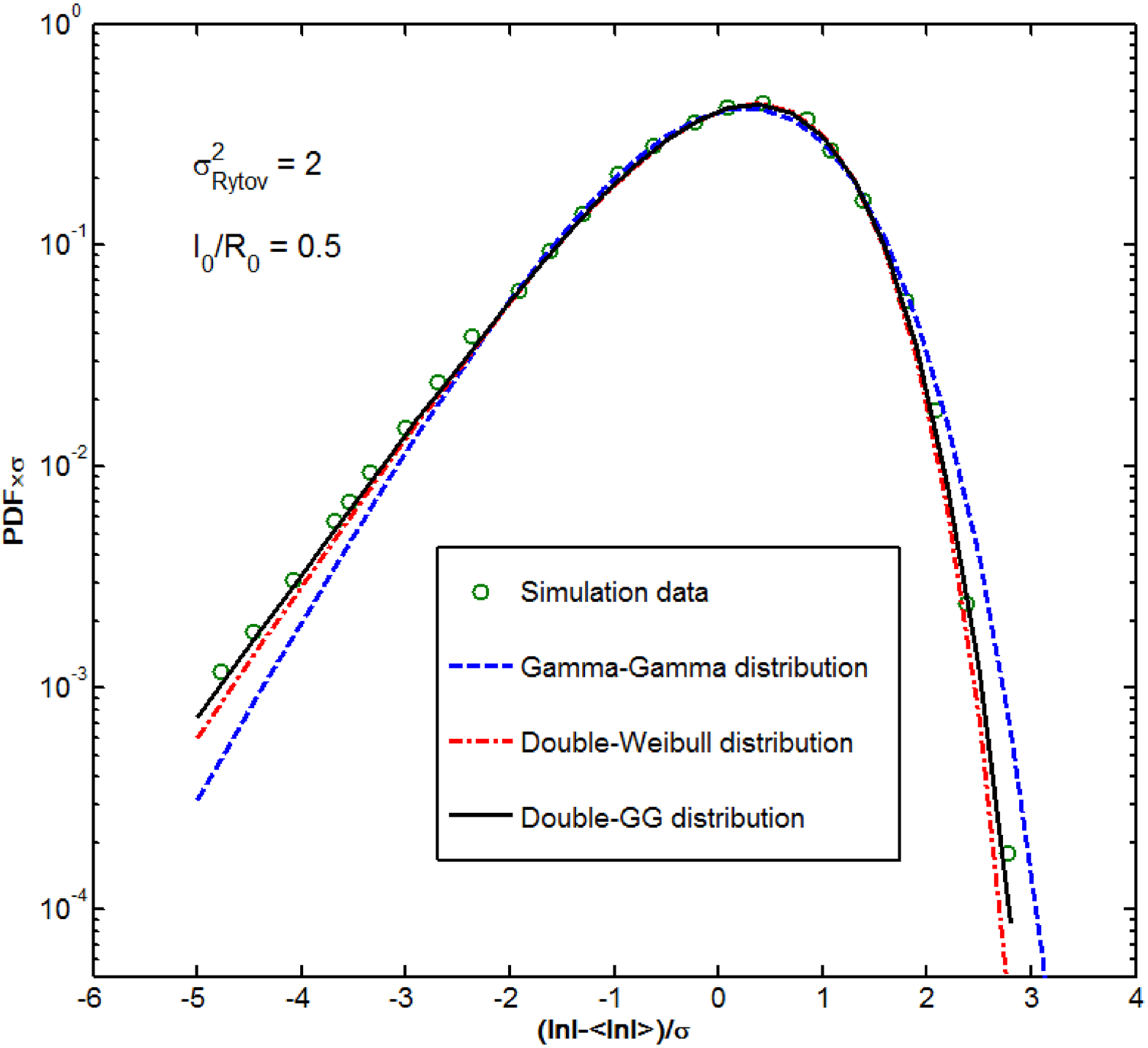}
\caption{Pdfs of the scaled log-irradiance for a plane wave assuming moderate irradiance fluctuations.}
\end{figure}

In the presence of a ﬁnite inner scale, the small-scale scintillation is again described by (\ref{R4}) and the large-scale variance is given by [\citen{3}, Eq. 78]
\begin{align}\nonumber
& \sigma _{x}^{2}\cong \exp \left[ 0.04\beta _{\text{0}}^{2}{{\left( \frac{8.56{{\eta }_{l}}}{8.56+{{\eta }_{l}}+0.195\beta _{\text{0}}^{2}\eta _{l}^{7/6}} \right)}^{7/6}} \right. \\\nonumber
& \times\left( 1+1.753{{\left( \frac{8.56}{8.56+{{\eta }_{l}}+0.195\beta _{\text{0}}^{2}\eta _{l}^{7/6}} \right)}^{1/2}}\right.\\\label{R7}
&\left.\left.-0.252{{\left( \frac{8.56}{8.56+{{\eta }_{l}}+0.195\beta _{\text{0}}^{2}\eta _{l}^{7/6}} \right)}^{7/12}} \right) \right]-1
\end{align}
Therefore, the parameters of the Double GG distribution are readily deduced from these expressions using only values of the refractive index structure parameter and inner scale according to the atmospheric conditions. The scintillation index can be further calculated as
\begin{equation}\label{eq10}
\sigma _{I}^{2}=\frac{E\left( {{I}^{2}} \right)}{E{{\left( I \right)}^{2}}}-1=\left( 1+\sigma _{x}^{2} \right)\left( 1+\sigma _{y}^{2} \right)-1
\end{equation}
We should emphasize that this distribution is very generic since it includes some commonly used fading models as special cases. For ${{\gamma }_{i}}\to 0$, ${{m}_{i}}\to \infty $, Double GG pdf coincides with the log-normal pdf. For ${{\gamma }_{i}}=1$, ${{\Omega }_{i}}=1$, it reduces to Gamma-Gamma while for ${{m}_{i}}=1$, it becomes Double-Weibull. For ${{\gamma }_{i}}=1$, ${{\Omega }_{i}}=1$, ${{m}_{2}}=1$, it coincides with the K channel.

\begin{figure}
\centering
\includegraphics[width = 8cm, height = 7.5cm]{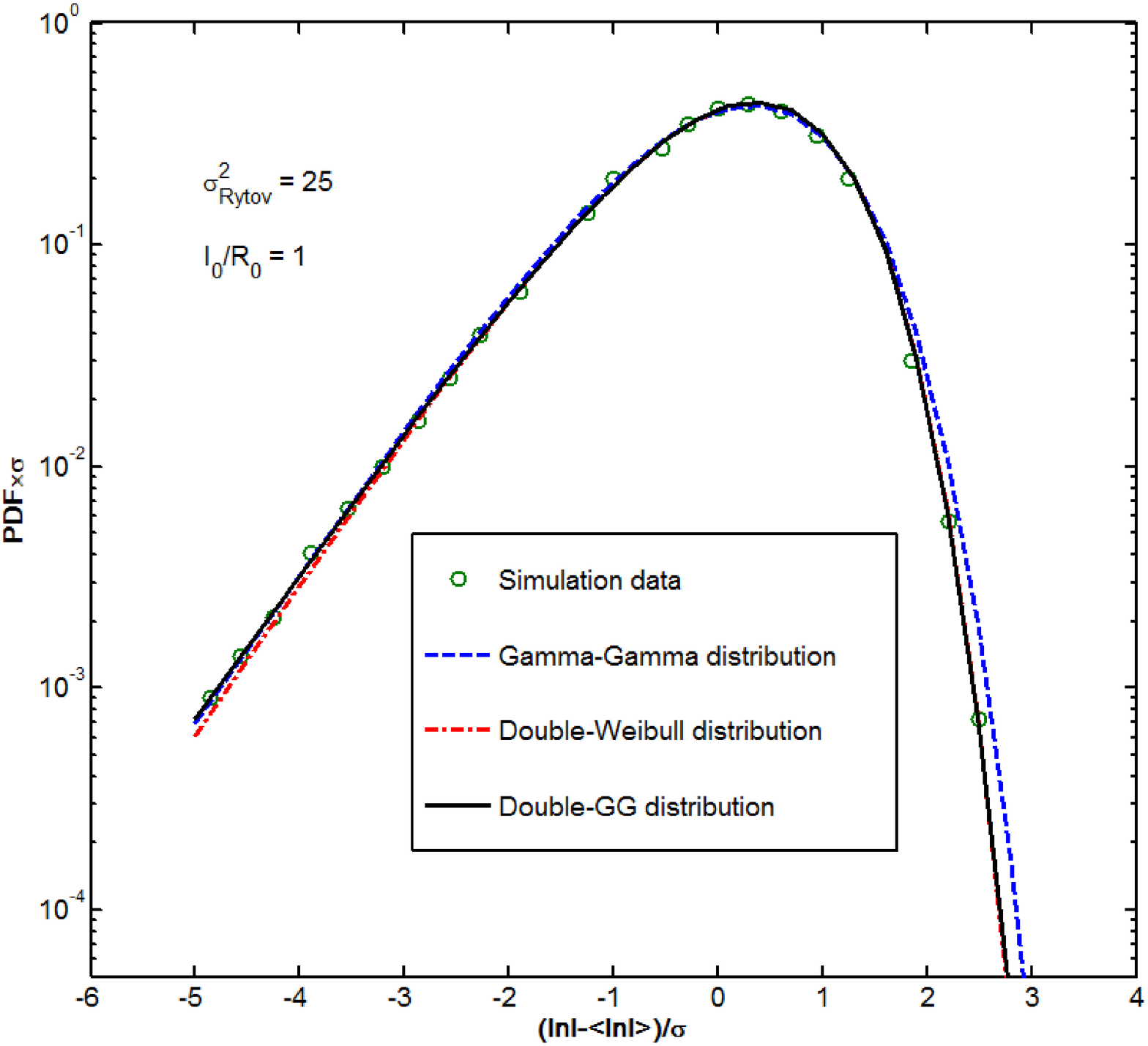}
\caption{Pdfs of the scaled log-irradiance for a plane wave assuming strong irradiance fluctuations.}
\end{figure}
\begin{table}
\begin{center}
\begin{quote}
\caption{NRMSE for different statistical models and turbulence conditions defined in Figs. 1-6}
\label{table2}
\end{quote}
\resizebox{\columnwidth}{!}{
\begin{tabular}{|c|c|c|c|}
  \hline
   &Gamma-Gamma \cite{10}  &Double-Weibull\cite{12} &Double GG (Proposed)\\ \hline
  Fig. 1  &2\% &7\% &0.6\% \\ \hline
  Fig. 2  &2.8\% &1.2\%   &0.8\% \\ \hline
  Fig. 3  &1.2\% &1\% &0.8\% \\ \hline
  Fig. 4  &0.3\% &10\% &0.3\% \\ \hline
  Fig. 5  &19\% &8.7\% &1.5\% \\ \hline
  Fig. 6  &4.8\% &2.4\% &1.7\% \\ \hline
  \end{tabular}}
\end{center}
\end{table}
\section{Verification of the Proposed Channel Model}\label{ver}
In this section, we compare the Double GG distribution model with simulation data of plane and spherical waves provided respectively in \cite{16} and \cite{17}. In \cite{16}, Flatt{\'e} \emph{et al.} carried out exhaustive numerical simulations and published the results assuming plane wave propagation through homogeneous and isotropic Kolmogorov turbulence. In \cite{17}, Hill and Frehlich presented the simulation data for the propagation of a spherical wave through homogeneous and isotropic atmospheric turbulence. The turbulence severity is characterized by Rytov variance ($\sigma _{\text{Rytov}}^{2}$) which is proportional to the scintillation index \cite{3}. We emphasize that the same data set was also employed in \cite{10} and \cite{12} which have introduced the Gamma-Gamma and Double-Weibull fading models.
\begin{figure}
\centering
\includegraphics[width = 8cm, height = 7.5cm]{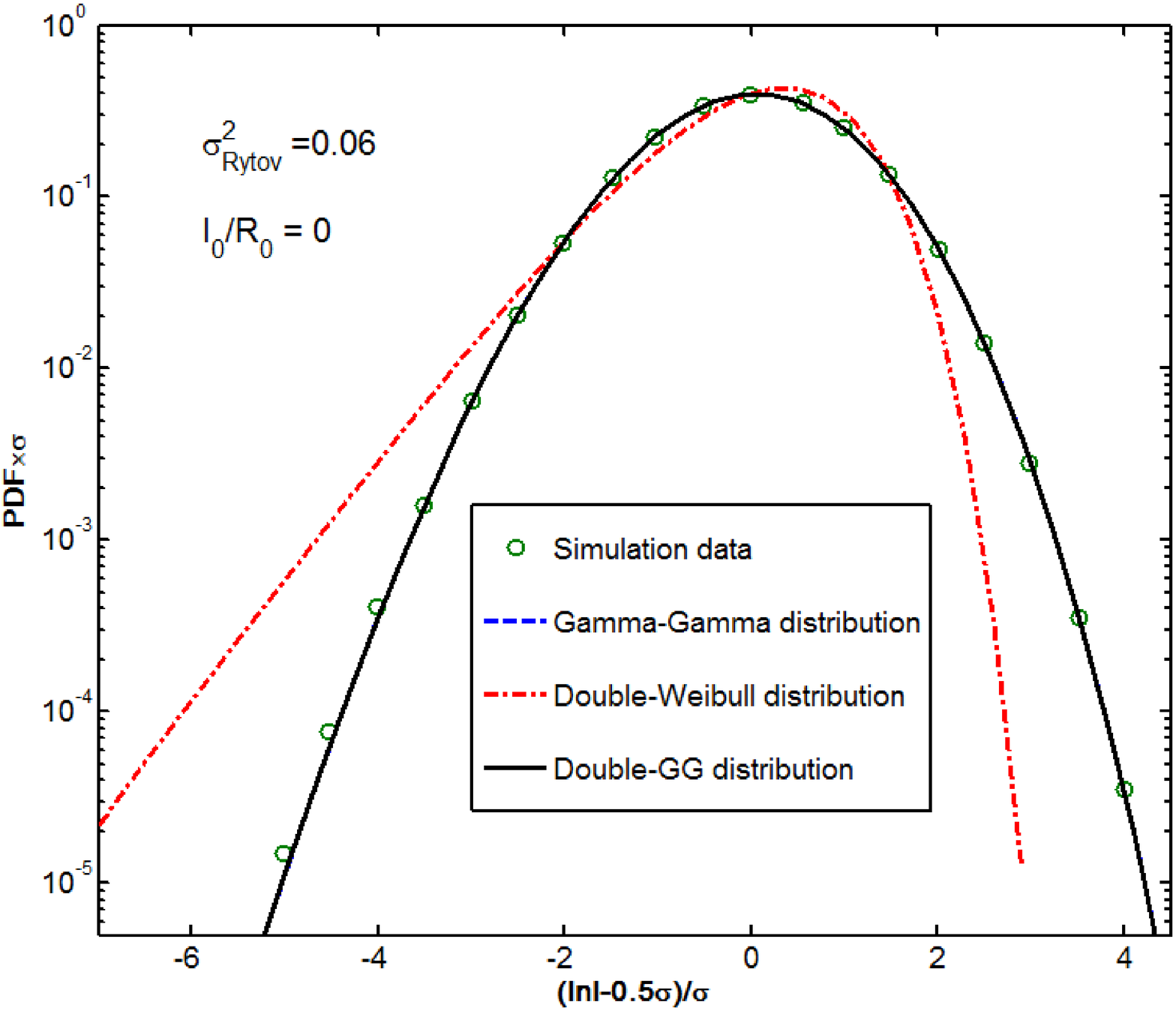}
\caption{Pdfs of the scaled log-irradiance for a spherical wave assuming weak irradiance fluctuations.}
\end{figure}
\subsection{Plane Wave}\label{plane}
Figs. 1-3 compare the Gamma-Gamma, Double-Weibull and Double GG models under a wide range of turbulence conditions (weak to strong) assuming plane wave propagation. In these figures, the vertical axis of the figure represents the log-irradiance pdf multiplied by the square root of variance. The logarithm of irradiance was particularly chosen to illustrate the high and low irradiance tails \cite{16}. Thus, sensitivity to the small irradiance fades is increased, while sensitivity to large irradiance peaks is decreased. The pdf plots were also scaled by subtracting the mean value to center all distributions on zero and dividing by the square root of variance.

In Fig. 1, we assume weak turbulence conditions which are characterized by $\sigma _{\text{Rytov}}^{2}=0.1$ and ${{l}_{0}}/{{R}_{0}}=0.5$. The values of the variances of small and large scale fluctuations, ($\sigma _{x}^{2}$ and $\sigma _{y}^{2}$) are calculated from (\ref{R1}) and (\ref{R2}). Using (\ref{eq8a}), (\ref{eq8b}) and (\ref{eq9}), the Double-GG parameters are obtained as ${{\gamma }_{1}}=2.1$, ${{\gamma }_{2}}=2.1$, ${{m}_{1}}=4$, ${{m}_{2}}=4.5$, ${{\Omega }_{1}}=1.0676$ and ${{\Omega }_{2}}=1.06$ where $p=q=1$. We further employ normalized root-mean-square error (NRMSE) as a statistical goodness of fit test. Table I provides the NRMSE results for different statistical models. According to Table I and Fig. 1, both Gamma-Gamma and Double-Weibull distributions fail to match the simulation data. On the other hand, the proposed Double GG distribution follows closely the simulation data.
\begin{figure}
\centering
\includegraphics[width = 8cm, height = 7.5cm]{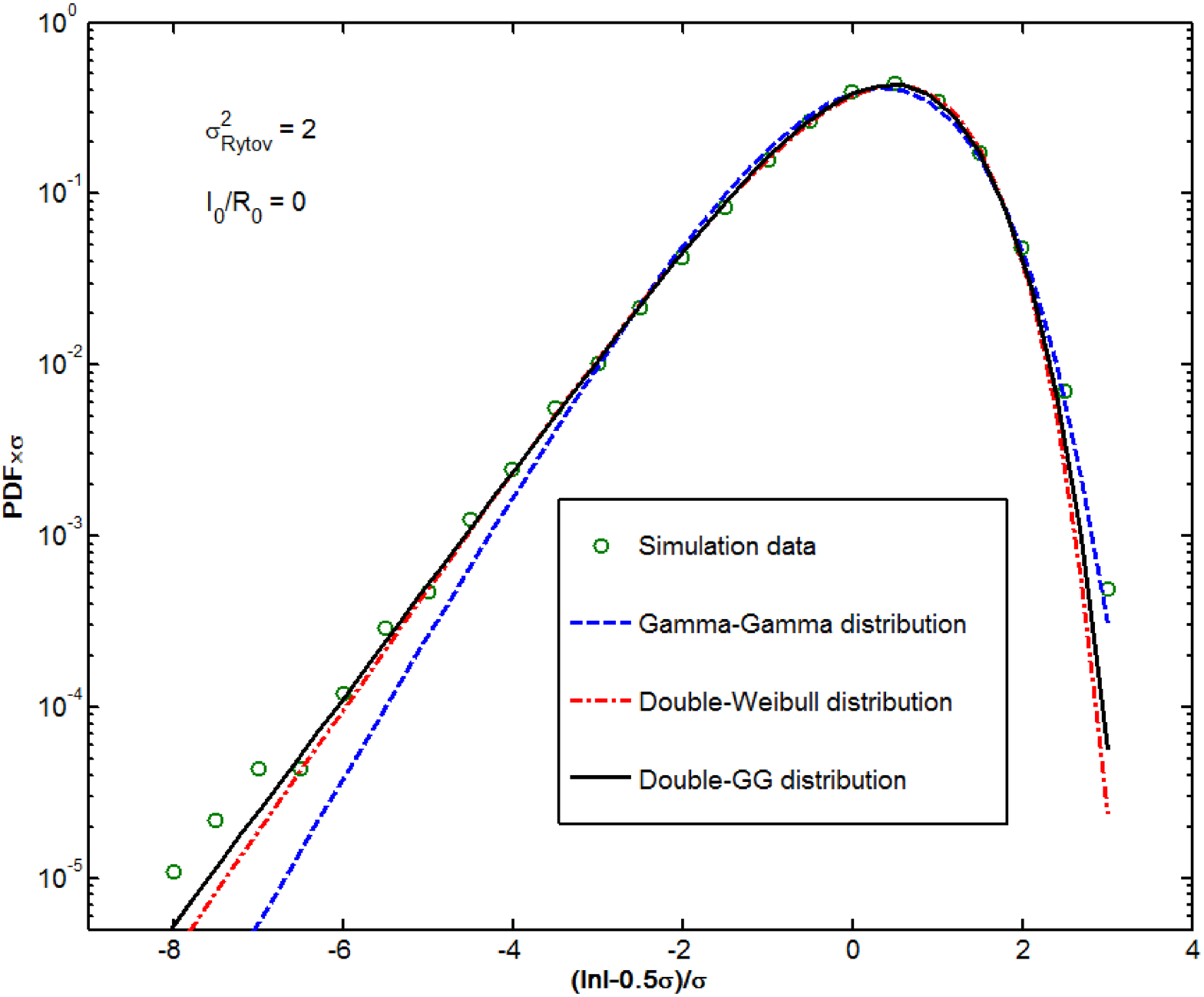}
\caption{Pdfs of the scaled log-irradiance for a spherical wave assuming moderate irradiance fluctuations.}
\end{figure}

In Fig.2, we assume moderate irradiance fluctuations which are characterized by $\sigma _{\text{Rytov}}^{2}=2$ and ${{l}_{0}}/{{R}_{0}}=0.5$. The parameters of the Double GG distribution for this case are obtained as ${{\gamma }_{1}}=2.1690$, ${{\gamma }_{2}}=0.8530$, ${{m}_{1}}=0.55$, ${{m}_{2}}=2.35$, ${{\Omega }_{1}}=1.5793$ and ${{\Omega }_{2}}=0.9671$. In the calculations, $p$ and $q$ are chosen as $p=28$ and $q=11$ in order to satisfy ${p}/{q}={{{\gamma }_{1}}}/{{{\gamma }_{2}}}$. Among the three distributions under consideration, the proposed Double GG model provides the best fit to the simulation data. It is apparent that Gamma-Gamma fails to match the simulation data particularly in the tails. As Table I demonstrates, the accuracy of Double Weibull is better than that of Gamma-Gamma, but slightly inferior to our proposed distribution.
\begin{figure}
\centering
\includegraphics[width = 8cm, height = 7.5cm]{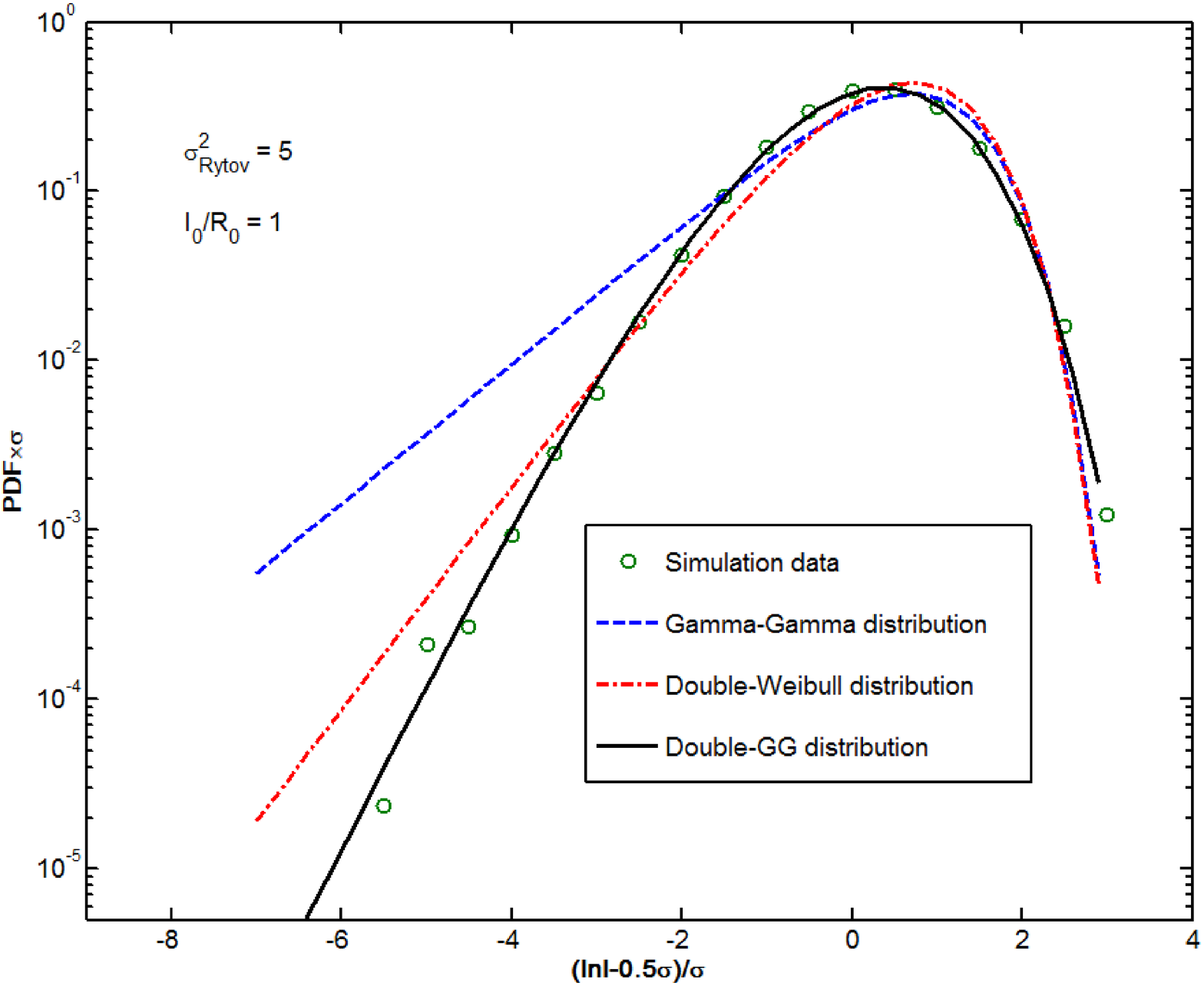}
\caption{Pdfs of the scaled log-irradiance for a spherical wave assuming strong irradiance fluctuations.}
\end{figure}

In Fig. 3, we assume strong irradiance fluctuations which are characterized by $\sigma _{\text{Rytove}}^{2}=25$ and ${{{I}_{0}}}/{{{R}_{0}}=1}$. The parameters of the Double GG distribution are calculated as ${{\gamma }_{1}}=1.8621$, ${{\gamma }_{2}}=0.7638$, ${{m}_{1}}=0.5$, ${{m}_{2}}=1.8$, ${{\Omega }_{1}}=1.5074$ and ${{\Omega }_{2}}=0.9280$ where $p$ and $q$ are chosen as 17 and 7 respectively. The Double GG model again provides an excellent match to the simulation data and as it is clear from Table I, its accuracy is better than Gamma-Gamma and Double Weibull.
\subsection{Spherical Wave}
\begin{figure} \centering
\subfigure[]{\includegraphics[width = 6.8cm, height = 5cm]{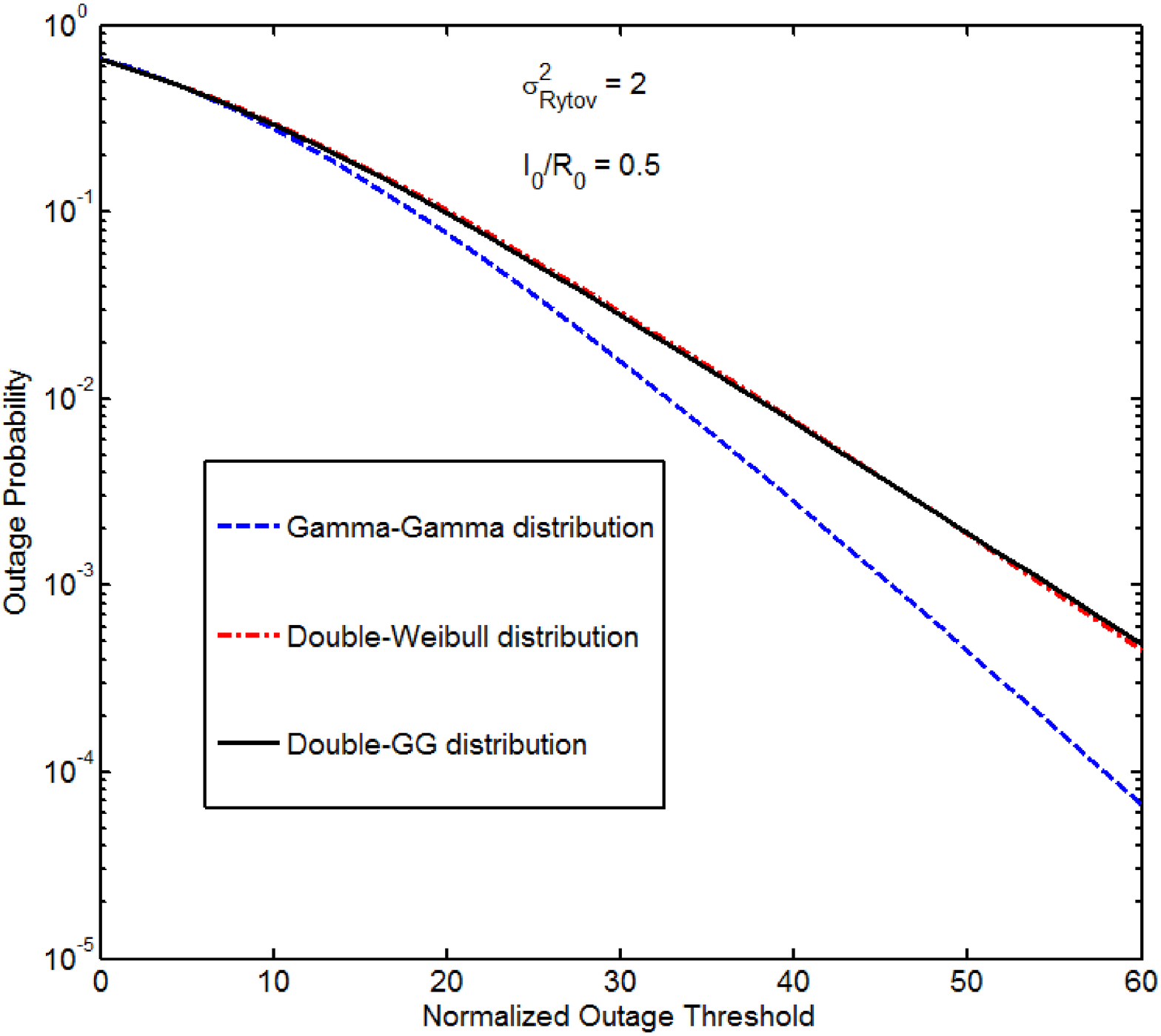}}
\subfigure[]{\includegraphics[width = 6.8cm, height = 5cm]{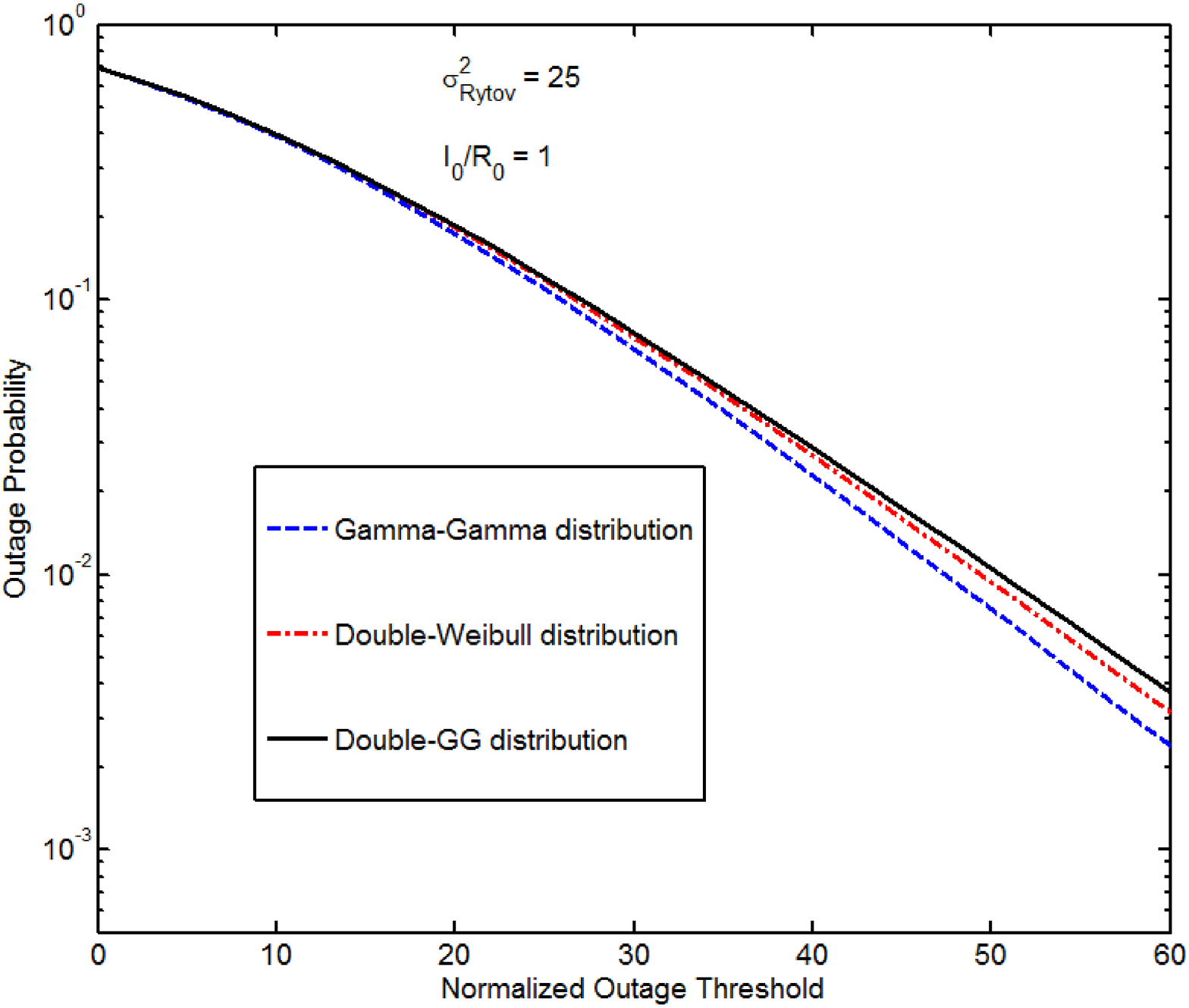}}
\subfigure[]{\includegraphics[width = 6.8cm, height = 5cm]{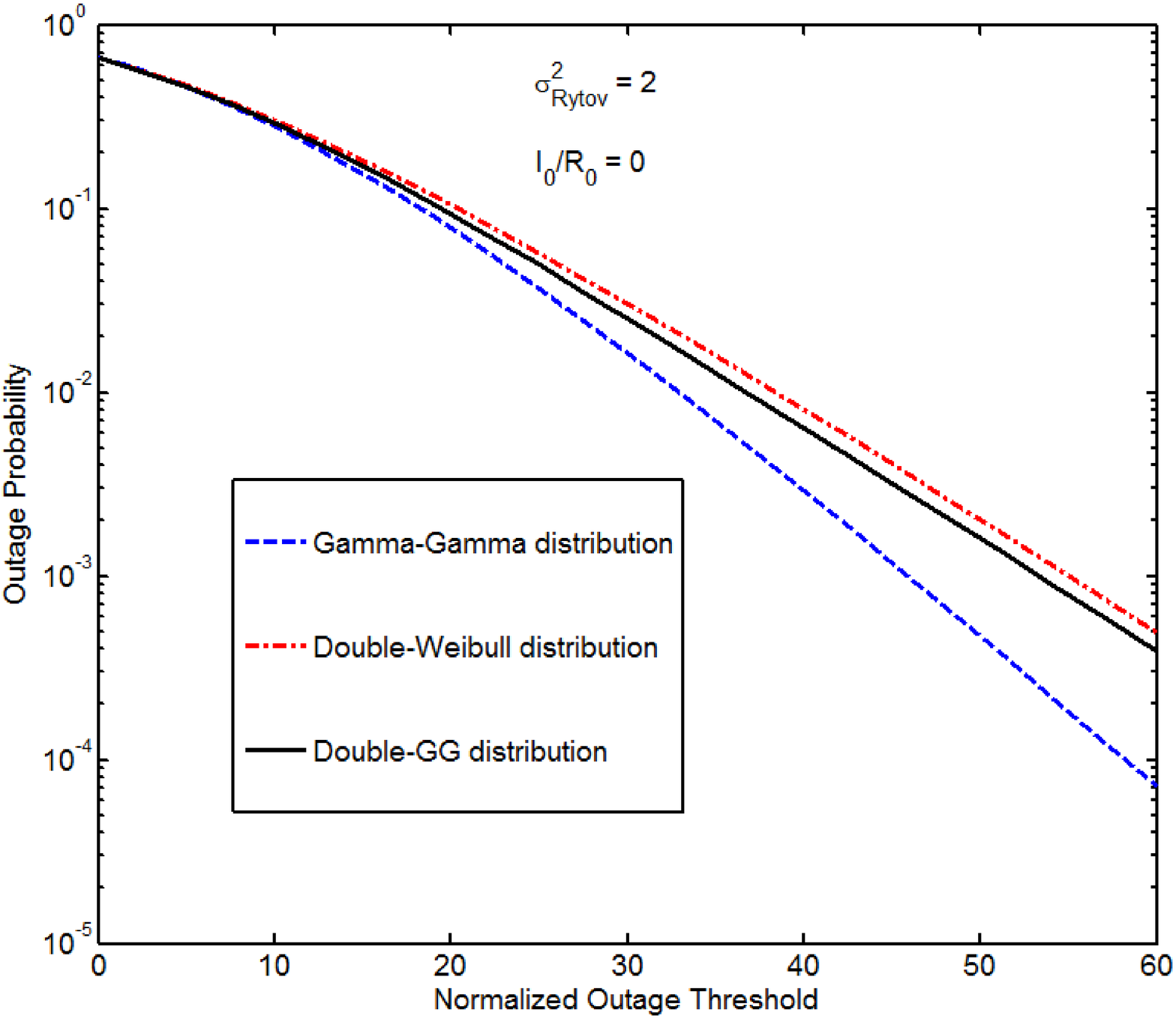}}
\subfigure[]{\includegraphics[width = 6.8cm, height = 5cm]{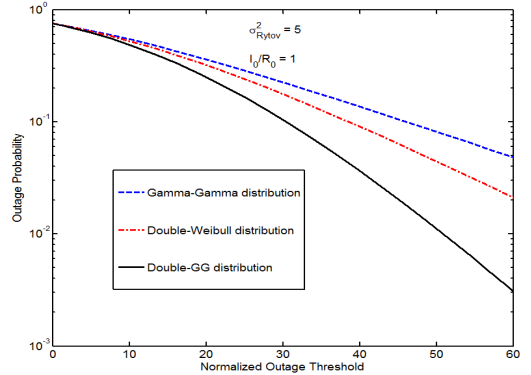}}
\caption{Outage probability as a function of $\bar{\gamma }/{{\gamma }_{th}}$  for a) Plane wave - $\sigma _{^{Rytov}}^{2}=2$, ${{{l}_{0}}}/{{{R}_{0}}=0.5}$, b) Plane wave-$\sigma _{^{Rytov}}^{2}=25$, ${{{l}_{0}}}/{{{R}_{0}}=1}$, c) Spherical wave - $\sigma _{^{Rytov}}^{2}=2$, ${{{l}_{0}}}/{{{R}_{0}}=0}$, d) Spherical wave - $\sigma _{^{Rytov}}^{2}=5$, ${{{l}_{0}}}/{{{R}_{0}}=1}$}
\end{figure}
Figs 4-6 compare the Gamma-Gamma, Double-Weibull and Double GG models under weak, moderate and strong turbulence conditions assuming spherical wave propagation. These pdfs are plotted as a function of $(\ln I+0.5{{\sigma }^{2}})/\sigma$ \cite{17}, where $\sigma$ is the square root of the variance of $\ln I$. The y-axis depicts the log-irradiance pdf multiplied by $\sigma$.

In Fig. 4, we consider spherical wave propagation and assume weak turbulence which are characterized by $\sigma _{\text{Rytov}}^{2}=0.06$ and ${{l}_{0}}/{{R}_{0}}=0$.  The parameters of Double GG are evaluated using the variances of small and large scale fluctuations, ($\sigma _{y}^{2}$ and $\sigma _{x}^{2}$) for spherical waves. The values of these variances are given by (\ref{R3}) , (\ref{R4}) and (\ref{R7}). Therefore, employing (9) and (\ref{eq9}) we obtain ${{m}_{1}}=34.24$, ${{m}_{2}}=32.79$, ${{\gamma }_{1}}={{\gamma }_{2}}={{\Omega }_{1}}={{\Omega }_{2}}=1$ where $p$ and $q$ are equal to 1. It can be noted that in this case, the Double GG coincides with the Gamma-Gamma distribution. It is apparent that both Gamma-Gamma and Double GG distributions provide an excellent match to the simulation data while the Double-Weibull distribution fails to match the simulation data.

In Fig. 5, we assume moderate irradiance fluctuations, which are characterized by $\sigma _{\text{Rytov}}^{2}=2$ and ${{{l}_{0}}}/{{{R}_{0}}=0}$. The parameters of the Double GG model for this case are calculated as ${{\gamma }_{1}}=0.9135$, ${{\gamma }_{2}}=1.4385$, ${{m}_{1}}=2.65$, ${{m}_{2}}=0.85$, ${{\Omega }_{1}}=0.9836$ and ${{\Omega }_{2}}=1.1745$ where $p$ and $q$ are selected as 7 and 11 respectively. It is clearly observed that the Double GG model provides a better fit with simulation data, especially for small irradiance values.

In Fig. 6, we assume strong irradiance fluctuations which are characterized by $\sigma _{\text{Rytov}}^{2}=5$ and ${{{l}_{0}}}/{{{R}_{0}}=1}$. The parameters of the Double GG model are calculated as ${{\gamma }_{1}}=0.4205$, ${{\gamma }_{2}}=0.6643$, ${{m}_{1}}=3.2$, ${{m}_{2}}=2.8$, ${\Omega_{1}}=0.8336$ and ${{\Omega }_{2}}=0.9224$ where $p$ and $q$ are chosen as 7 and 11 respectively. It is apparent from this figure and Table I that both Gamma-Gamma and Double-Weibull distributions fail to match the simulation data. On the other hand, the proposed Double GG distribution follows closely the simulation data.
\section{Performance Evaluation}
\subsection{Outage Probability Analysis of SISO FSO System}
Denote ${{R}_{t}}$ as a targeted transmission rate and assume ${{\gamma }_{th}}={{C}^{-1}}\left( {{R}_{t}} \right)$ as the corresponding signal-to-noise ratio (SNR) threshold in terms of the instantaneous channel capacity for a particular channel realization. Therefore, the outage probability is calculated by ${{P}_{out}}\left( {{R}_{t}} \right)=Pr\left( \gamma <{{\gamma }_{th}} \right)$ \cite{26,kashani}. If SNR exceeds $\gamma_{th}$, no outage happens and the receiver can decode the signal with arbitrarily low error probability. For the system under consideration, the instantaneous electrical SNR can be defined as $\gamma ={{\left( \eta I \right)}^{2}}/{{N}_{0}}$, while the average electrical SNR is obtained as $\bar{\gamma }={{{\eta }^{2}}}/{{{N}_{0}}}\;$ since $E\left( I \right)=1$. Therefore, $I$ can be expressed as $I=\sqrt{\gamma /\bar{\gamma }}$. After the transformation of the random variable, $I$, the cdf of $\gamma $ can be easily derived from (\ref{eq5}) and setting $\gamma ={{\gamma }_{th}}$ therein, we obtain the outage probability as in (\ref{eq11}) at the top of the next page.
\begin{figure*}[t]
\begin{equation}\label{eq11}
{{P}_{out}}={{F}_{\gamma }}\left( {{\gamma }_{th}} \right)
=\frac{{{p}^{{{m}_{2}}-1/2}}{{q}^{{{m}_{1}}-1/2}}{{\left( 2\pi  \right)}^{{1-\left( p+q \right)}/{2}\;}}}{\Gamma \left( {{m}_{1}} \right)\Gamma \left( {{m}_{2}} \right)}
G_{1,p+q+1}^{p+q,1}\left[ {{\left( \frac{{{\gamma }_{th}}}{{\bar{\gamma }}} \right)}^{p{{\gamma }_{2}}/2}}\frac{m_{1}^{q}m_{2}^{p}}{{{\left( {{\Omega }_{2}}p \right)}^{p}}{{\left( {{\Omega }_{1}}q \right)}^{q}}}|\begin{matrix}
   1  \\
   \Delta \left( q:{{m}_{1}} \right),\Delta \left( p:{{m}_{2}} \right),0  \\
\end{matrix} \right]
\end{equation}
\hrulefill
\end{figure*}

(\ref{eq11}) can be seen as a generalization of earlier outage analysis results in the literature. Specifically, if we insert ${{\gamma }_{i}}=1$ and ${{\Omega }_{i}}=1$ in (\ref{eq11}), we obtain the outage probability expression reported in [\citen{18}, Eq. (15)] for Gamma-Gamma channel.  Setting ${{m}_{i}}=1$ in (\ref{eq11}), we recover Eq (16) of \cite{12} derived for Double-Weibull channel. Similarly, for ${{\gamma }_{i}}=1$, ${{\Omega }_{i}}=1$ and ${{m}_{2}}=1$, (\ref{eq11}) reduces to (3) of \cite{27} reported for the K channel.
\begin{figure} \centering
\subfigure[]{\includegraphics[width = 6.8cm, height = 5cm]{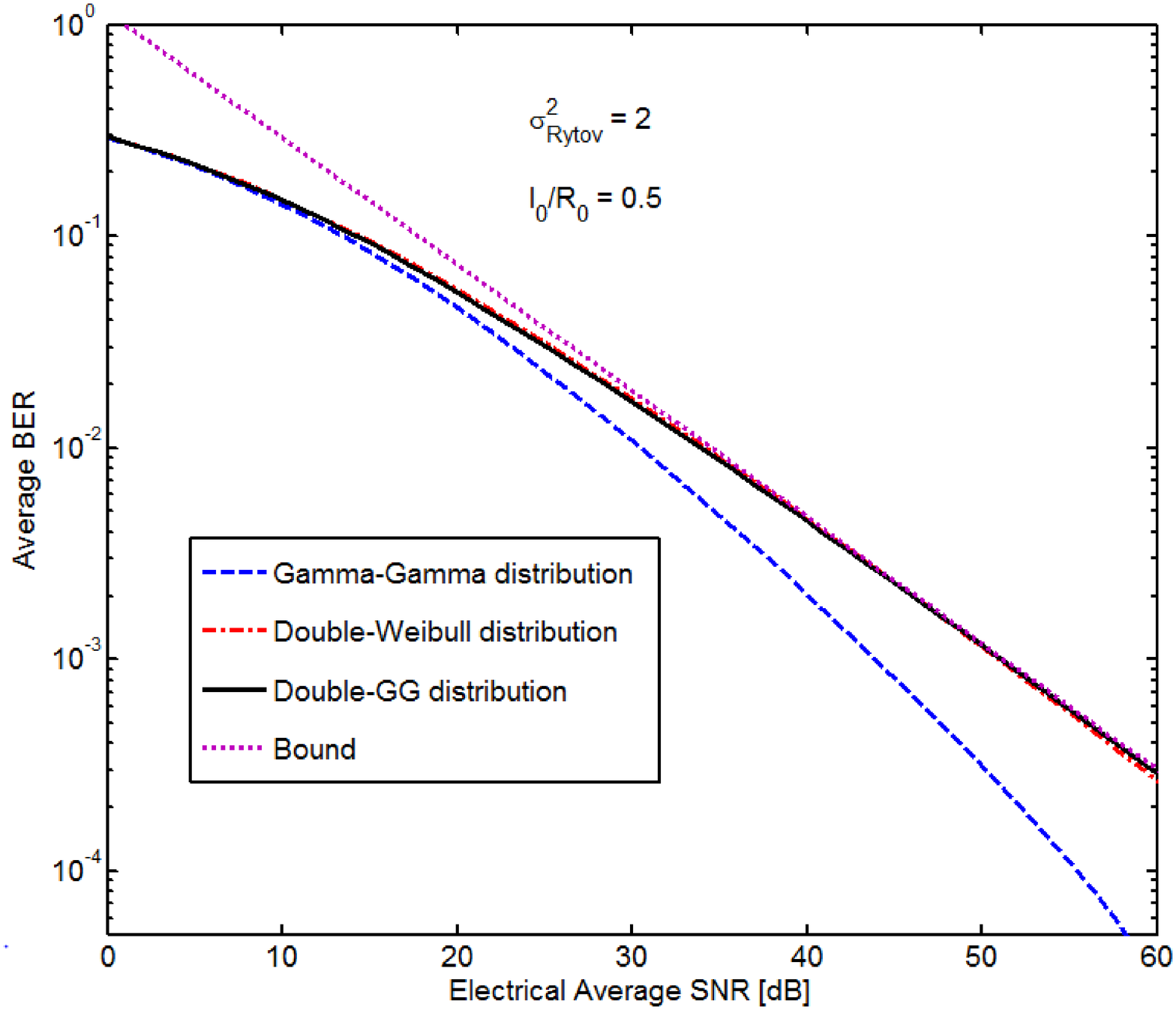}}
\subfigure[]{\includegraphics[width = 6.8cm, height = 5cm]{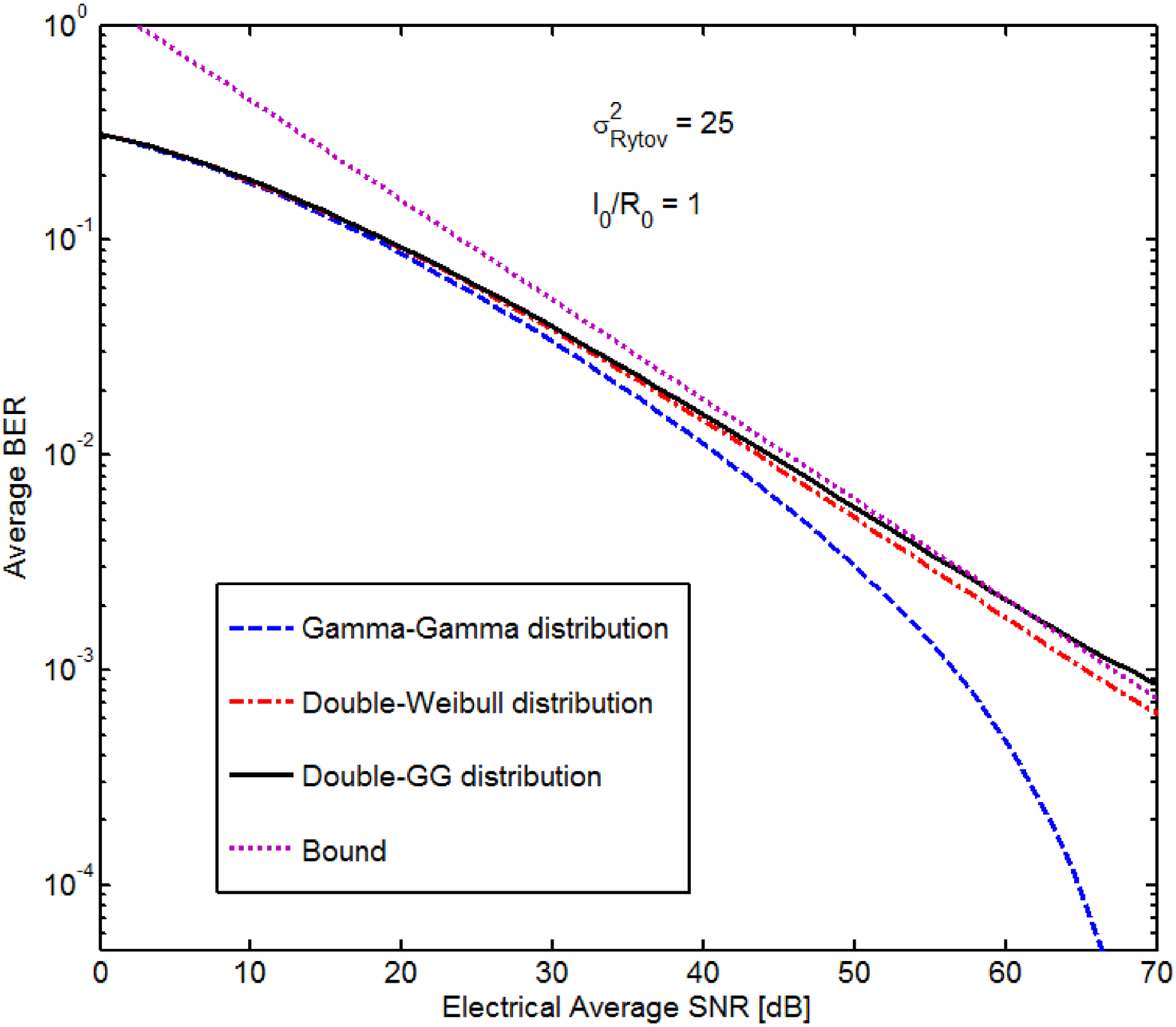}}
\subfigure[]{\includegraphics[width = 6.8cm, height = 5cm]{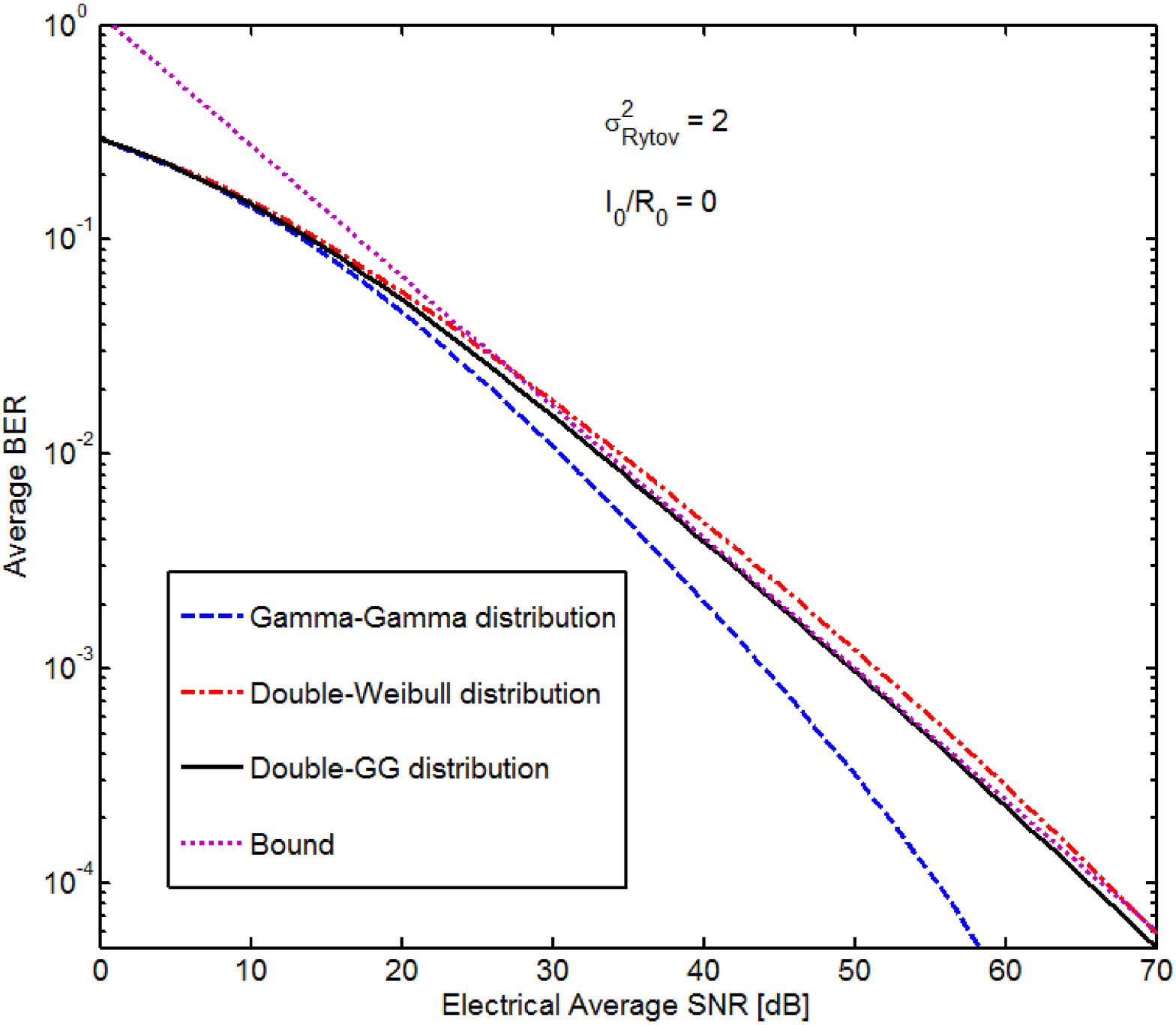}}
\subfigure[]{\includegraphics[width = 6.8cm, height = 5cm]{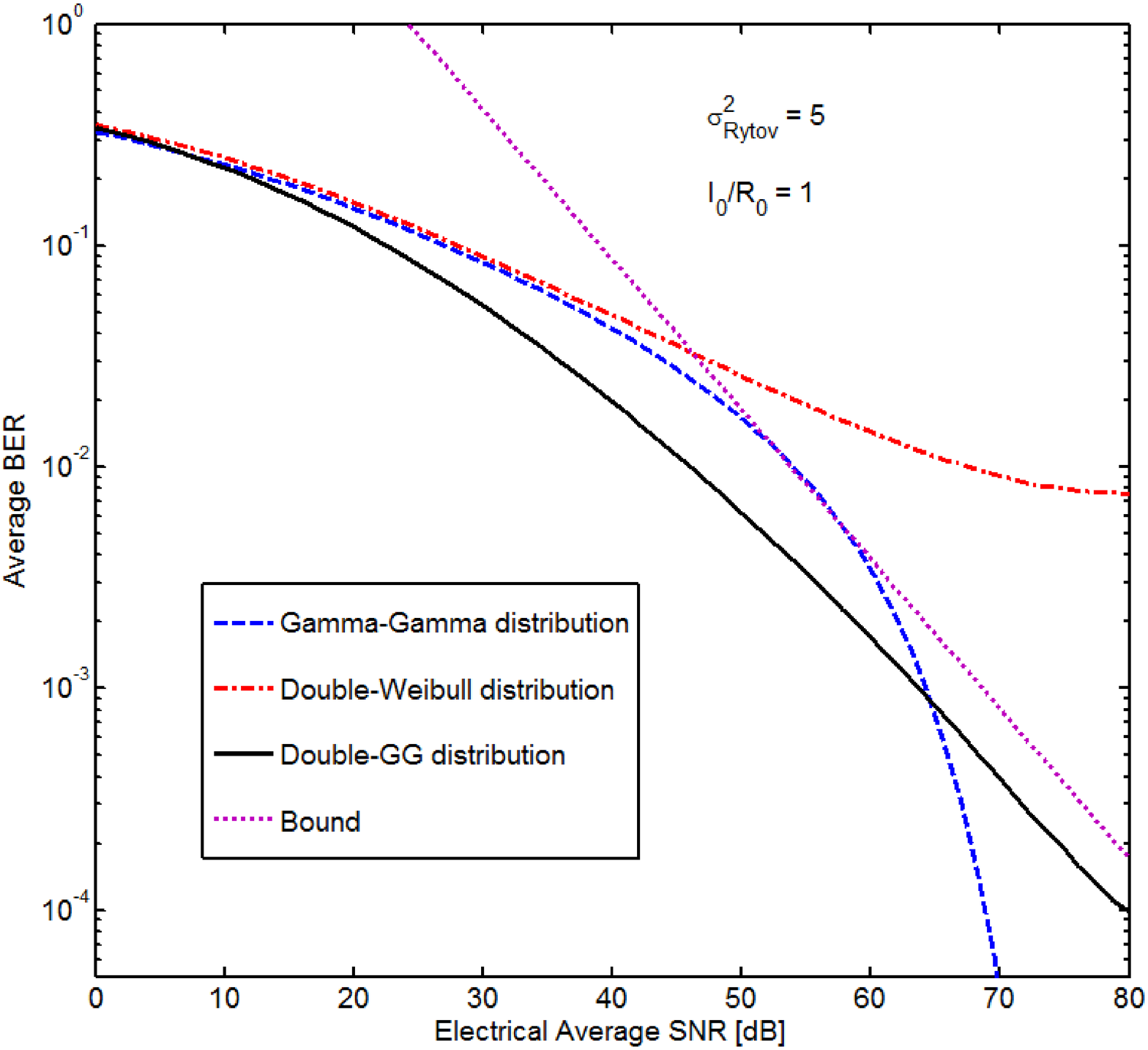}}
\caption{Average BER as a function of $\bar{\gamma }$ a) Plane wave - $\sigma _{^{Rytov}}^{2}=2$, ${{{l}_{0}}}/{{{R}_{0}}=0.5}$, b) Plane wave-$\sigma _{^{Rytov}}^{2}=25$, ${{{l}_{0}}}/{{{R}_{0}}=1}$, c) Spherical wave - $\sigma _{^{Rytov}}^{2}=2$, ${{{l}_{0}}}/{{{R}_{0}}=0}$, d) Spherical wave - $\sigma _{^{Rytov}}^{2}=5$, ${{{l}_{0}}}/{{{R}_{0}}=1}$}
\end{figure}

Based on the derived expression in (\ref{eq11}), Figs. 7.a-d present the outage probabilities of a SISO FSO system for different degrees of turbulence severity. We adopt the same parameters used in Figs 2-3 and 5-6 and consider the following four cases: a) Plane wave with $\sigma _{\text{Rytov}}^{2}=2$ and ${{{l}_{0}}}/{{{R}_{0}}=0.5}$, b) Plane wave with $\sigma _{\text{Rytov}}^{2}=25$ and ${{{l}_{0}}}/{{{R}_{0}}=1}$, c) Spherical wave with $\sigma _{\text{Rytov}}^{2}=2$ and ${{{l}_{0}}}/{{{R}_{0}}=0}$, d) Spherical wave with $\sigma _{\text{Rytov}}^{2}=5$  and ${{{l}_{0}}}/{{{R}_{0}}=1}\;$. It is observed from Fig. 7.a that an SNR of 37.8 dB is required to achieve a targeted outage probability of $10^{-2}$. As the turbulence strength gets stronger (see Fig. 7.b), the required SNR to maintain the same performance climbs up to 50.5 dB. Similarly, for spherical waves, SNRs of 36.8 dB and 50.9 dB are respectively required for moderate and strong turbulence conditions. In these figures, we further include the outage results for Double-Weibull and Gamma-Gamma for comparison purposes. As expected from the earlier comparisons of their pdfs, the outage performance over Double-Weibull and Double-GG for plane wave (See Figs. 7.a and 7.b) are similar while the Gamma-Gamma model overestimates the outage performance. On the other hand, the superiority of Double GG is more obvious for spherical wave (See Figs 7.c and 7.d), particularly for strong turbulence conditions, where the outage performance plots of Double-Weibull and Gamma Gamma significantly deviate.
\subsection{BER Analysis of SISO FSO System}
In this section, we present the BER performance analysis of an FSO system with on-off keying (OOK) over the proposed Double GG channel. The received optical signal is written as
\begin{equation}\label{eq12}
y=\eta Ix+n
\end{equation}
where $x$ represents the information bits and can be either 0 or 1, $n$ is the Additive White Gaussian noise (AWGN) term with zero mean and variance $\sigma _{n}^{2}={{N}_{0}}/2$ , $\eta$ is the optical-to-electrical conversion coefficient and $I$ is the normalized irradiance whose pdf follows (\ref{eq4}). Conditioned on the irradiance, the instantaneous BER for OOK is given by \cite{20}
\begin{equation}\label{eq13}
{{P}_{e,ins}}=0.5\operatorname{erfc}\left( \frac{\eta I}{2\sqrt{{{N}_{0}}}} \right)
\end{equation}
where $\operatorname{erfc}\left( . \right)$ stands for the complementary error function defined as
\begin{equation}\label{eq14}
\operatorname{erfc}(x)=\frac{2}{\sqrt{\pi }}\int_{x}^{\infty }{{{e}^{-{{t}^{2}}}}}dt.
\end{equation}
The average BER can be then calculated by averaging (\ref{eq13}) over the distribution of $I$, i.e.,
\begin{equation}\label{eq15}
{{P}_{e}}=\int_{0}^{\infty }{{{f}_{I}}\left( I \right)}\left[ 0.5\operatorname{erfc}\left( \frac{\eta I}{2\sqrt{{{N}_{0}}}} \right) \right]dI
\end{equation}
The above integral can be evaluated in closed form by expressing the $\operatorname{erfc}\left( . \right)$ integrand via a Meijer’s G-function using [\citen{28}, Eq. (8.4.14.2)], [\citen{28}, Eq. (8.2.2.14)] and [\citen{29}, Eq. (21)]. Thus, a closed-form solution is obtained as in (\ref{eq16}) at the top of the next page.

\begin{figure*}[t]
  \begin{equation}\label{eq16}
   {{P}_{SISO}}=\frac{{{\gamma }_{2}}{{k}^{{{m}_{1}}+{{m}_{2}}}}{{p}^{{{m}_{2}}+1/2}}{{q}^{{{m}_{1}}-1/2}}}{{{2}^{\frac{3}{2}}}l\Gamma \left( {{m}_{1}} \right)\Gamma \left( {{m}_{2}} \right){{\left( 2\pi  \right)}^{\frac{l+k\left( p+q \right)}{2}-1}}}
  G_{2l,k\left( p+q \right)+l}^{k\left( p+q \right),2l}\left[ {{\left( \frac{m_{1}^{q}m_{2}^{p}}{{{p}^{P}}\Omega _{2}^{p}{{q}^{q}}\Omega _{1}^{q}} \right)}^{k}}\frac{{{\left( 4l \right)}^{l}}}{{{{\bar{\gamma }}}^{l}}{{k}^{k\left( p+q \right)}}}|\begin{matrix}
   \Delta \left( l:1 \right),\Delta \left( l:\frac{1}{2} \right)  \\
   {{\operatorname{J}}_{k}}\left( q:1-{{m}_{1}} \right),{{\operatorname{J}}_{k}}\left( p:1-{{m}_{2}} \right),\Delta \left( l:0 \right)  \\
\end{matrix} \right]
\end{equation}
\hrulefill
\end{figure*}

In (\ref{eq16}) $k$ and $l$ are positive integer numbers that satisfy ${p{{\gamma }_{2}}}/{2}\;={l}/{k}\;$ and ${{\operatorname{J}}_{\xi }}\left( y,x \right)$ is defined as
\begin{align}\label{eq17}
&{{\operatorname{J}}_{\xi }}\left( y,x \right)\\\nonumber
&=\Delta \left( \xi ,\frac{y-x}{y} \right),\Delta \left( \xi ,\frac{y-1-x}{y} \right),\ldots ,\Delta \left( \xi ,\frac{1-x}{y} \right)
\end{align}

The derived BER expression in (\ref{eq16}) can be seen as a generalization of earlier BER results in the literature. If we insert ${{\gamma }_{i}}=1$ and ${{\Omega }_{i}}=1$ in (\ref{eq16}), we obtain the BER expression derived in [\citen{19}, Eq. (9)] under the assumption of Gamma-Gamma channel. Setting ${{m}_{i}}=1$ in (\ref{eq16}), we obtain Eq (15) of \cite{12} derived for Double-Weibull channel. On the other hand, for ${{\gamma }_{i}}=1$, ${{\Omega }_{i}}=1$ and ${{m}_{2}}=1$, (\ref{eq16}) reduces to (12) of \cite{20} reported for the K-channel.

In an effort to have some further insights into system performance, we investigate the asymptotical BER performance in the following. For large SNR values, the asymptotic BER behavior is dominated by the behavior of the pdf near the origin, i.e. ${{f}_{I}}\left( I \right)$ at $I\to 0$ \cite{garcia}. Thus, employing series expansion corresponding to the Meijer’s G-function [\citen{wol}, Eq. (07.34.06.0006.01)], the Double GG distribution given in (\ref{eq4}) can be approximated by a single polynomial term as
\begin{equation}\label{app}
{{f}_{I}}\left( I \right)\approx A \prod\limits_{\begin{smallmatrix}
 j=1 \\
 j\ne k
\end{smallmatrix}}^{p+q}{\Gamma \left( {{b}_{j}}-{{b}_{k}} \right)}{{I}^{p{{\gamma }_{2}}\min \left\{ \frac{{{m}_{1}}}{q},\frac{{{m}_{2}}}{p} \right\}-1}}
\end{equation}
where $A$ is obtained as
\begin{align}\nonumber
&A =\frac{{{\gamma }_{2}}p{{p}^{{{m}_{2}}-1/2}}{{q}^{{{m}_{1}}-1/2}}{{\left( 2\pi  \right)}^{{1-\left( p+q \right)}/{2}\;}}}{\Gamma \left( {{m}_{1}} \right)\Gamma \left( {{m}_{2}} \right)}\\\label{aa}
&\times{{\left( \frac{m_{1}^{q}m_{2}^{p}}{{{\left( q{{\Omega }_{1}} \right)}^{q}}{{\left( p{{\Omega }_{2}} \right)}^{p}}} \right)}^{\min \left\{ \frac{{{m}_{1}}}{q},\frac{{{m}_{2}}}{p} \right\}}}
\end{align}
In (\ref{app}), ${{b}_{k}}$ and ${{b}_{j}}$ are defined as
\begin{equation}
{{b}_{k}}=\min \left\{ \frac{{{m}_{1}}}{q},\frac{{{m}_{2}}}{p} \right\}
\end{equation}
\begin{align}
&{{b}_{j}}\in \\\nonumber
&\left\{ 1-\Delta \left( q:1-{{m}_{1}} \right),1-\Delta \left( p:1-{{m}_{2}} \right) \right\}\backslash \min \left\{ \frac{{m}_{1}}{q},\frac{{m}_{2}}{p} \right\}
\end{align}
Therefore, based on (\ref{eq15}), the average BER can be well approximated by
\begin{equation}\label{siso_sym}
{{P}_{SISO}}\approx A \prod\limits_{\begin{smallmatrix}
 j=1 \\
 j\ne k
\end{smallmatrix}}^{p+q}{\Gamma \left( {{b}_{j}}-{{b}_{k}} \right)}{{\left( \frac{2}{\sqrt{{\bar{\gamma }}}} \right)}^{p{{\gamma }_{2}}{{b}_{k}}}}\frac{\Gamma \left( \left( 1+p{{\gamma }_{2}}{{b}_{k}} \right)/2 \right)}{2\sqrt{\pi }p{{\gamma }_{2}}{{b}_{k}}}
\end{equation}
From (\ref{siso_sym}), it can be readily deduced that the diversity order of SISO FSO system is given by $0.5p{{\gamma }_{2}}\min \left\{ {{{m}_{1}}}/{q}\;,{{{m}_{2}}}/{p}\right\}$.
\begin{figure}
\centering
\includegraphics[width = 8cm, height = 7.5cm]{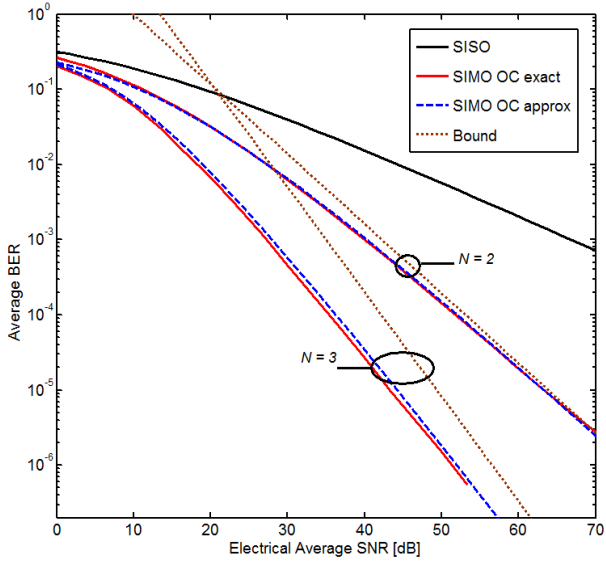}
\caption{Comparison of the average BER between SISO and SIMO with optimal combing for plane wave as defined in Fig. 3.}
\end{figure}

It is observed from Fig.8.a that an SNR of 51.1 dB is required to achieve a BER of $10^{-3}$ for a plane wave in moderate turbulence conditions. For stronger turbulence conditions, the required SNR to achieve the same BER performance is 68.2 dB as seen from Fig. 8.b. For spherical waves, SNRs of 49.8 dB and 63.8 are respectively required for moderate and strong turbulence conditions. Comparison with the expressions presented for other channel models reveals that the Gamma-Gamma model significantly overestimates the performance. Similar to earlier observations on the outage analysis, the superiority of Double GG is more obvious for spherical wave. As observed from Figs 8.c and 8.d, the performance plots of Double-Weibull and Gamma Gamma considerably diverge particularly for strong turbulence conditions.  Furthermore, it can be clearly seen that the asymptotic results are in excellent agreement with exact analytical results within a wide range of SNR showing the accuracy and usefulness of the derived asymptotic expression given in (\ref{siso_sym}).
\section{FSO Links with Receive Diversity}\label{div}
\begin{figure}
\centering
\includegraphics[width = 8cm, height = 7.5cm]{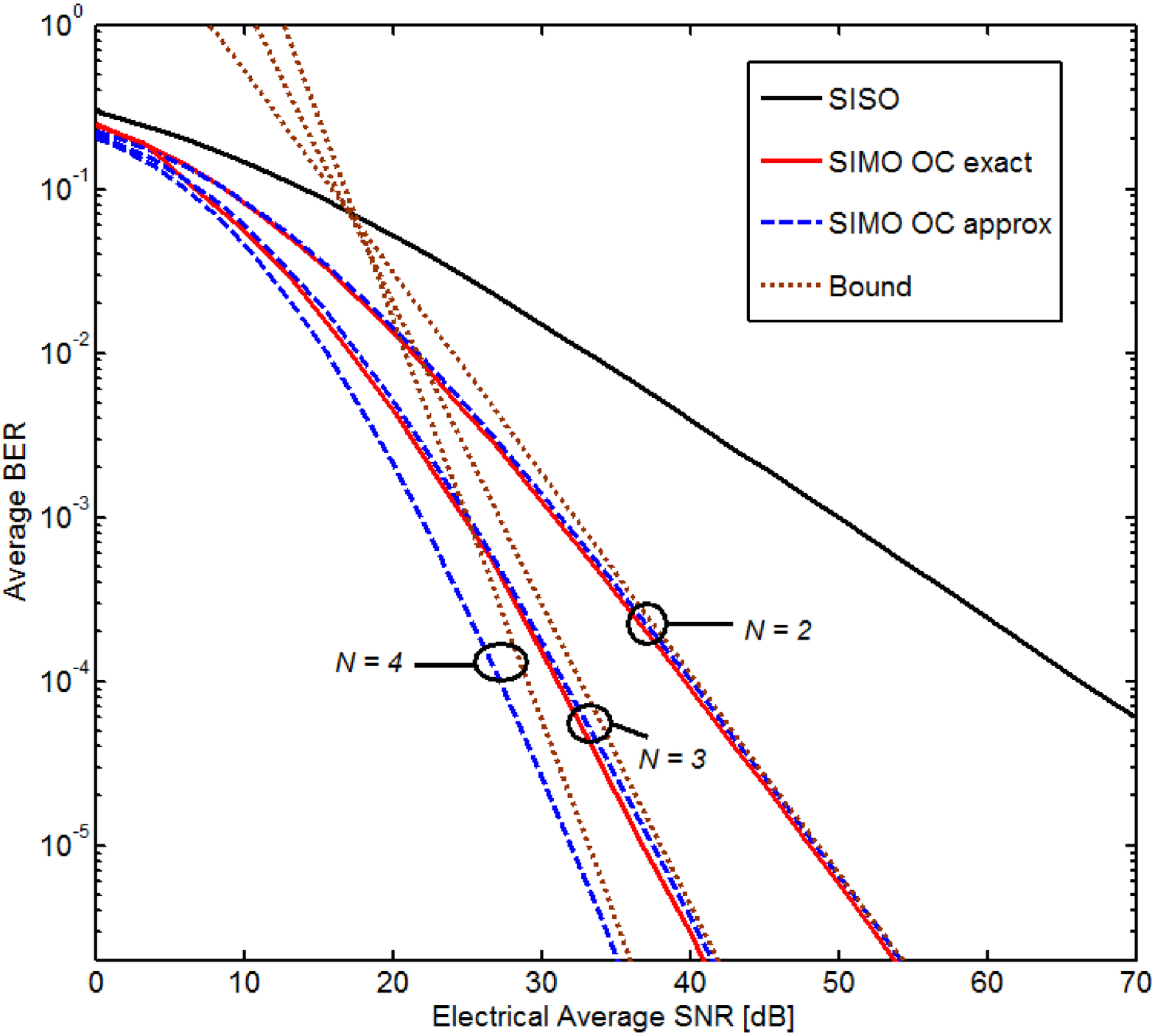}
\caption{Comparison of the average BER between SISO and SIMO with optimal combing for spherical wave as defined in Fig. 5.}
\end{figure}
As it can be noticed from Section IV, the performance of a SISO FSO link over moderate and strong atmospheric turbulence is quite poor. To address this issue, multiple transmit and/or receive apertures can be employed and the performance can be improved via diversity gains. In the following, we assume that multiple receive apertures are employed and present the BER derivations under the assumption that optimal gain combining is used.

The optimum decision metric for OOK is given by \cite{21}
\begin{equation}\label{eq18}
P\left( \mathbf{r}|\text{on,}{{\text{I}}_{n}} \right)\underset{\text{off}}{\overset{\text{on}}{\mathop{\lessgtr }}}\,P\left( \mathbf{r}|\text{off,}{{\text{I}}_{n}} \right)
\end{equation}
where $\mathbf{r}=\left( {{r}_{1}},{{r}_{2}},...,{{r}_{N}} \right)$ is the received signal vector and ${{I}_{n}}$ is the fading channel coefﬁcient which models the channel from the transmit aperture to the $n^{\text{th}}$ receive aperture. Following the same approach as \cite{20,21} the conditional bit error probabilities are given by
\begin{equation}\label{eq19}
{{P}_{e}}(\text{off }|{{I}_{n}})={{P}_{e}}(\text{on }|{{I}_{n}})=Q\left( \frac{1}{N}\sqrt{\frac{{\bar{\gamma }}}{2}\sum\limits_{n=1}^{N}{I_{n}^{2}}} \right)
\end{equation}
The average BER can be then obtained as
\begin{equation}\label{eq20}
{{P}_{\text{SIMO,OC}}}=\int\limits_{\mathbf{I}}{{{f}_{\mathbf{I}}}}\left( \mathbf{I} \right)Q\left( \sqrt{\frac{{\bar{\gamma }}}{2N}\sum\limits_{n=1}^{N}{I_{n}^{2}}} \right)d\mathbf{I}
\end{equation}
where ${{f}_{\mathbf{I}}}\left( \mathbf{I} \right)$ is the joint pdf of vector $\mathbf{I}=\left( {{I}_{1}},{{I}_{2}},\ldots ,{{I}_{N}} \right)$. The factor $N$ in (\ref{eq20}) is used to ensure that sum of the $N$ receive aperture areas is the same as the area of the receive aperture of the SISO link.
\begin{figure*}[t]
\begin{equation}\label{eq24}
\Lambda \left( n,\upsilon  \right)=\frac{{{\alpha }_{n}}l_{n}^{-0.5}k_{n}^{{{m}_{1,n}}+{{m}_{2,n}}}}{2{{\left( 2\pi  \right)}^{0.5\left( {{l}_{n}}-1+\left( {{k}_{n}}-1 \right)\left( {{p}_{n}}+{{q}_{n}} \right) \right)}}}G_{{{l}_{n}},{{k}_{n}}\left( {{p}_{n}}+{{q}_{n}} \right)}^{{{k}_{n}}\left( {{p}_{n}}+{{q}_{n}} \right),{{l}_{n}}}\left[ \frac{{{\left( \upsilon N \right)}^{{{l}_{n}}}}\omega _{n}^{-{{k}_{n}}}l_{n}^{{{l}_{n}}}}{{{{\bar{\gamma }}}^{{{l}_{n}}}}k_{n}^{{{k}_{n}}\left( {{p}_{n}}+{{q}_{n}} \right)}}\left| \begin{matrix}
   \Delta \left( {{l}_{n}},1 \right)  \\
   {{J}_{{{k}_{n}}}}\left( {{q}_{n}},1-{{m}_{1,n}} \right),{{J}_{{{k}_{n}}}}\left( {{p}_{n}},1-{{m}_{2,n}} \right)  \\
\end{matrix} \right. \right]
\end{equation}
\hrulefill
\end{figure*}

Eq. (\ref{eq20}) does not yield a closed-form solution and requires N-dimensional integration. Nevertheless, the Q-function can be well-approximated as $Q(x)\approx {{{e}^{-\frac{{{x}^{2}}}{2}}}}/{12}\;+{{{e}^{-\frac{2{{x}^{2}}}{3}}}}/{4}\;$ \cite{30} and thus the average BER can be obtained as
\begin{align}\nonumber
{{P}_{\text{SIMO,OC}}}&\approx \frac{1}{12}\prod\limits_{n=1}^{N}{\int_{0}^{\infty }{{{f}_{{{I}_{n}}}}\left( {{I}_{n}} \right)}}\exp \left( \frac{-\bar{\gamma }}{4N}I_{n}^{2} \right)d{{I}_{n}}\\\label{eq21}
&+\frac{1}{4}\prod\limits_{n=1}^{N}{\int_{0}^{\infty }{{{f}_{{{I}_{n}}}}\left( {{I}_{n}} \right)}}\exp \left( \frac{-\bar{\gamma }}{3N}I_{n}^{2} \right)d{{I}_{n}}
\end{align}
The above integral can be evaluated by first expressing the exponential function in terms of the Meijer G-function presented in [\citen{29}, eq. (11)] as
\begin{equation}\label{eq22}
\exp \left( -x \right)=\operatorname{G}_{0,1}^{1,0}\left[ x\left| _{0}^{-} \right. \right]
\end{equation}
Then, a closed-form expression for (\ref{eq21}) is obtained using [\citen{29}, Eq. (21)] as
\begin{equation}\label{eq23}
{{P}_{\operatorname{SIMO},OC}}\approx \frac{1}{12}\prod\limits_{n=1}^{N}{\Lambda \left( n,4 \right)}+\frac{1}{4}\prod\limits_{n=1}^{N}{\Lambda \left( n,3 \right)}
\end{equation}
where $\Lambda \left( n,\upsilon  \right)$ is defined in (\ref{eq24}) at the top of the page.

In (\ref{eq24}), ${{l}_{n}}$ and ${{k}_{n}}$ are positive integer numbers that satisfy ${{{p}_{n}}{{\gamma }_{2,n}}}/{2}\;={{{l}_{n}}}/{{{k}_{n}}}\;$, and ${{\alpha }_{n}}$ and ${{\omega }_{n}}$ $n\in \left\{ 1,2,\ldots ,N \right\}$ are defined as
\begin{align}\label{eq25}
&{{\alpha }_{n}}=\frac{{{\gamma }_{2,n}}p_{n}^{{{m}_{2,n}}+1/2}q_{n}^{{{m}_{1,n}}-1/2}{{\left( 2\pi  \right)}^{{1-\left( {{p}_{n}}+{{q}_{n}} \right)}/{2}\;}}}{\Gamma \left( {{m}_{1,n}} \right)\Gamma \left( {{m}_{2,n}} \right)}\\\label{eq25b}
&{{\omega }_{n}}={{\left( {{\Omega }_{2,n}}{{p}_{n}}m_{2,n}^{-1} \right)}^{{{p}_{n}}}}{{\left( {{q}_{n}}{{\Omega }_{1,n}}m_{1,n}^{-1} \right)}^{{{q}_{n}}}}
\end{align}
The derived expression in (\ref{eq23}) includes the previously reported result in \cite{20} for K channel as a special case.

Based on the approximation in (\ref{app}), the corresponding closed-form asymptotic solution for (\ref{eq21}) can be obtained as
\begin{equation}\label{simo_asy}
{{P}_{\operatorname{SIMO},OC\_\text{asy}}}\approx \frac{1}{12}\prod\limits_{n=1}^{N}{{{\Lambda }_{\text{asy}}}\left( n,4 \right)}+\frac{1}{4}\prod\limits_{n=1}^{N}{{{\Lambda }_{\text{asy}}}\left( n,3 \right)}
\end{equation}
where ${{\Lambda }_{\text{asy}}}\left( n,\upsilon  \right)$ is defined as
\begin{align}\nonumber
&{{\Lambda }_{\text{asy}}}\left( n,\upsilon  \right)={{\alpha }_{n}}\prod\limits_{\begin{smallmatrix}
 j=1 \\
 j\ne k
\end{smallmatrix}}^{{{p}_{n}}+{{q}_{n}}}{\Gamma \left( {{b}_{j,n}}-{{b}_{k,n}} \right)}\\
&\times\Gamma \left( {{p}_{n}}{{\gamma }_{2,n}}{{b}_{k,n}} \right)\frac{{{\left( \sqrt{\upsilon N} \right)}^{{{p}_{n}}{{\gamma }_{2,n}}{{b}_{k,n}}}}}{2{{\left( \sqrt{{\bar{\gamma }}} \right)}^{{{p}_{n}}{{\gamma }_{2,n}}{{b}_{k,n}}}}}
\end{align}
Therefore, the diversity order of FSO links with $N$ receive apertures employing optimal gain combining is obtained as $0.5\sum\limits_{n=1}^{N}{{{p}_{n}}{{\gamma }_{2,n}}\min \left\{ {{{m}_{1}}_{,n}}/{{{q}_{n}}},{{{m}_{2,n}}}/{{{p}_{n}}}\right\}}$.

Figs. 9-10 illustrate the BER performance of the SIMO FSO system under consideration. We present approximate analytical results which have been obtained through (\ref{eq23}) and (\ref{simo_asy}) along with the Monte-Carlo simulation of (\ref{eq20}). As clearly seen from Figs. 9-10, our approximate expressions provide an excellent match to simulation results. As a benchmark, the BER of SISO FSO link is also included in these figures. It is observed that receive diversity signiﬁcantly improve the performance. For instance, at a target bit error rate of ${{10}^{-3}}$, we observe performance improvements of 26.8 dB and 39.6 dB respectively for with $N=2$ and 3 with respect to the SISO transmission over Double GG turbulence channels defined in Fig.3. Similarly, for Double GG channels defined in Fig. 5, at a BER of ${{10}^{-3}}$, performance improvements of 19 dB and 25.1 dB are achieved for SIMO links with $N=2$ and 3 compared to the SISO deployment. It can be further observed that asymptotic bounds on the BER become tighter at high enough SNRs confirming the accuracy and usefulness of the asymptotic expression given in (\ref{simo_asy}).

\section{Conclusions}\label{Con}
In this paper, we have introduced a new channel model, so called Double GG, which accurately describes irradiance fluctuations over atmospheric channels under a wide range of turbulence conditions. It is based on the theory of doubly stochastic scintillation and considers irradiance fluctuations as the product of small-scale and large-scale fluctuations which are both Generalized Gamma distributed. We have obtained closed-form expressions for the pdf and cdf in terms of Meijer’s G-function. Comparisons with the Gamma Gamma and Double-Weibull have shown that the new model provides an accurate fit with numerical simulation data for both plane and spherical waves. Using the new channel model, we have obtained closed-form expressions for the BER and the outage probability of SISO and SIMO FSO systems. We have demonstrated that our derived expressions cover many existing results in the literature earlier reported for Gamma-Gamma, Double-Weibull and K channels as special cases. Based on the asymptotical performance analysis, we have further derived diversity gains for SISO and SIMO FSO systems under consideration.

\balance

\bibliographystyle{IEEEtr}
\bibliography{DGG_2}
\end{document}